\documentclass[12pt]{iopart}

\usepackage{iopams}
\usepackage{mathrsfs}
\usepackage{latexsym}
\usepackage{subfigure, graphicx}
\usepackage{caption}
\begin{document}
	
	\title[Deep Transfer Learning for aLIGO Detector Characterization]{Deep Transfer Learning: A new deep learning glitch classification method for advanced LIGO}

	\author{Daniel George$^{1, 2}$, Hongyu Shen$^{1,3}$ and E.~A. Huerta$^1$}
	\address{$^1$ NCSA, University of Illinois at Urbana-Champaign, Urbana, Illinois, 61801}
	\address{$^2$ Department of Astronomy, University of Illinois at Urbana-Champaign, Urbana, Illinois, 61801}
	\address{$^3$ Department of Statistics, University of Illinois at Urbana-Champaign, Urbana, Illinois, 61801}
	
	\vspace{10pt}
	\begin{indented}
		\item[\today]
	\end{indented}
	
	\begin{abstract}
		
		The exquisite sensitivity of the advanced LIGO detectors has enabled the detection of multiple gravitational wave signals, thereby establishing gravitational wave astrophysics as an active field of research. The sophisticated design of these detectors mitigates the effect of most types of noise. However, advanced LIGO data streams are contaminated by numerous artifacts known as glitches --- non-Gaussian noise transients with complex morphologies. Given their high rate of occurrence, glitches can lead to false coincident detections, obscure and even mimic true gravitational wave signals. Therefore, successfully characterizing and removing glitches from advanced LIGO data is of utmost importance. In this article, we present the first application of \textit{Deep Transfer Learning} for glitch classification, showing that knowledge from deep learning algorithms trained for real-world object recognition can be transferred for classifying glitches in time-series based on their spectrogram images. Using the \texttt{Gravity Spy} dataset, containing hand-labeled, multi-duration spectrograms obtained from real LIGO data, we demonstrate that this method enables optimal use of very deep convolutional neural networks for classification given small training datasets, significantly reduces the time for training the networks, and achieves state-of-the-art accuracy above 98.8\%, with perfect precision-recall on 8 out of 22 classes. Furthermore, new types of glitches can be classified accurately given few labeled examples with this technique. Once trained via transfer learning, we show that the convolutional neural networks can be truncated and used as excellent feature extractors for unsupervised clustering methods to identify new classes based on their morphology, without any labeled examples. Therefore, this provides a new framework for dynamic glitch classification for gravitational wave detectors, which are expected to encounter new types of noises as they undergo gradual improvements to attain design sensitivity.
	\end{abstract}
	
	%
	\vspace{2pc}

	%
	%
	%
	%

	\section{Introduction}
	\label{intro}
	
	The advanced Laser Interferometer Gravitational wave Observatory (aLIGO) detectors are the largest and most sensitive interferometric detectors ever built. The cutting-edge design and fabrication of aLIGO's subsystems enable the sensing of changes in aLIGO's arm-length thousands of times smaller than the diameter of a proton~\cite{LSC:2015,DII:2016}. This instrument has already detected multiple gravitational wave (GW) signals produced from mergers of black holes~\cite{GW1,GW2,thirddetection}. As aLIGO gradually attains design sensitivity, it will transition from its current discovery mode into an astronomical observatory that will routinely detect new GW sources, providing insights into astrophysical objects and processes that cannot be observed through any other means~\cite{LVCT,SathyaLRR:2009}. 
	
	
	The study of glitches is of paramount importance to characterize GW detectors. For aLIGO to realize its full potential, it is necessary to ensure that its sensing capabilities are not hindered by unwanted noises that contaminate GW data. This is a non-trivial task requiring ``intelligent'' algorithms given that noise transients vary widely in duration, frequency range and morphology, spanning a wide parameter space that is challenging to model accurately~\cite{corn:2015CQGra,jade:2015CQGra,jade1:2016,DBNN}. Furthermore, since the aLIGO detectors are undergoing commissioning between each observing run, we expect that new types of glitches will be identified as aLIGO attains design sensitivity~\cite{DII:2016,D7:2016,D8:2016}. Accurately characterizing glitches is essential to find and eliminate noise transients that can obscure GW signals, or accidentally lead to coincident false detections which mimic GW signals, due to their high occurrence rate.   
	
	The complex and time-evolving nature of glitches makes them an ideal case study to apply machine learning algorithms --- methods that learn from examples rather than being explicitly programmed. Machine learning can be divided into supervised and unsupervised learning depending on whether labeled, structured data is used for training the algorithms~\cite{DL-Book}. Deep learning, i.e, machine learning based on deep artificial neural networks~\cite{DL-Nature, DL-Book}, is one of the fastest growing fields of artificial intelligence research today, having outperformed competing methods in many areas of machine learning applications, e.g., image classification, face recognition, natural language understanding and translation, speech recognition and synthesis, game-playing (e.g., Go, Poker), and self-driving vehicles. Therefore, deep learning algorithms are also expected to have excellent performance for the characterization of gravitational wave detectors. In this article, we focus on deep learning with Convolutional Neural Networks (CNNs), the leading approach for computer visions tasks~\cite{DL-Nature}, using spectrograms computed from the time-series glitches as inputs.
	
	Recent efforts on this front include \texttt{Gravity Spy}, an innovative interdisciplinary project that provides an infrastructure for citizen scientists to label datasets of glitches from aLIGO via crowd-sourcing~\cite{spy:2016arXiv}. Supervised classification algorithms based on this dataset were presented in the first Gravity Spy article~\cite{spy:2016arXiv} and were further discussed in Ref.~\cite{multi:2017arXiv}. In the latter study, deep multi-view models were introduced to enhance classification accuracy using multiple-duration spectrograms of each glitch stitched-together to form a larger input image. These algorithms employed deep learning CNN models which were 4 layers deep, and achieved overall accuracies close to 97\% for glitch classification. However, it was found that glitch classes with very few labeled samples were more difficult to classify with the same level of accuracy.
	
	As new classes of glitches are uncovered in the near future, the design and hyperparameters~\footnote{Hyperparameters refer to several quantities that have to be manually chosen to determine the architecture of the neural networks (e.g., overall design, number of layers, sizes of the convolutions, depth, padding, types of layers and activations, etc.)} of these CNNs will have to be modified accordingly to distinguish subtle features between a larger set of classes. Considering these issues, one may be interested in exploring the use of very deep CNNs designed for the harder task of classifying thousands of categories of real-world objects. However, the small size of the \texttt{Gravity Spy} dataset, compared to the datasets used to train these state-of-the-art CNNs for real-world object recognition, would lead to poor results due to immediate overfitting, i.e., memorization of features in the training data without generalizing to the testing data. Therefore, previous attempts at glitch classification from spectrograms have used \textit{relatively} shallow CNNs, which were designed and trained from the ground-up~\cite{spy:2016arXiv, multi:2017arXiv}. With this approach, lower input image resolution is used to keep model size manageable while significant time and effort is needed to optimize the architecture (hyperparameters) of the CNNs and to train them from scratch. To directly employ deep CNN models with well-established architectures optimized for computer vision, for this glitch classification problem, a different approach is necessary.
	
	In this article, we present \textit{Deep Transfer Learning}, a new method for glitch classification that leverages the complex structure and deep abstraction of pre-trained state-of-the-art CNNs used for object recognition and fine-tunes them throughout all layers to accurately classify glitches after re-training on a small dataset of aLIGO spectrograms. We show that this technique achieves state-of-the-art results for glitch classification with the \texttt{Gravity Spy} dataset, attaining above 98.8\% overall accuracy and perfect precision-recall on 8 out of 22 classes, while significantly reducing the training time to a few minutes. We also demonstrate that features learned from real-world images by very deep CNNs are directly transferable for the classification of spectrograms of time-series data from GW detectors, and possibly also for spectrograms in general, although the two datasets are very dissimilar. The CNNs we use were originally designed for over 1000 classes of objects in ImageNet. Therefore, our algorithm can be easily extended to classify hundreds of new classes of glitches in the future, especially since this transfer learning approach requires only a few labeled examples of a new class. Furthermore, we outline how new classes of glitches can be automatically found and grouped together as they occur using our trained CNNs as feature extractors for unsupervised or semi-supervised clustering algorithms.

	This article is organized as follows. Section~\ref{met} describes the methods and datasets used for applying transfer learning to develop accurate CNNs for glitch classification. Section~\ref{exp} summarizes our results. We discuss immediate applications of our method and its extension for finding new classes of glitches in an unsupervised manner in Section~\ref{Dis}. Conclusions and a brief description of future strategies to assist ongoing efforts for aLIGO data analysis are provided in Section~\ref{end}.

	\section{Methods}
	\label{met}
	
	In this Section we provide a succinct description of neural networks and transfer learning. Thereafter, we present our methods and procedure for applying transfer learning in the context of glitch classification.
	
	\subsection{Neural Networks}
	
	CNNs are currently the best performing method for image classification, object recognition, and a variety of other image processing tasks~\cite{DL-Nature}. They serve as excellent feature extractors by allowing end-to-end learning of features and representations from raw image data for classification, thus eliminating the need for feature engineering, i.e., hand-extraction of features or representations by human experts. CNN-based algorithms have recently achieved super-human results in the ImageNet Large Scale Visual Recognition Competition (ILSVRC)~\cite{ImageNet}, and has been outperforming all other methods since 2012~\cite{NIPS2012_4824}.
	
	State-of-the-art CNNs for real-world object recognition are extremely powerful, and thus training them from the ground up on a small dataset would lead to poor performance. This is because the capacity of the model, i.e., the degrees of freedom (number of parameters) of the model would be much larger than necessary to fit the small data distribution, thus resulting in what is known as ``overfitting''(see section 5.2 in the Deep Learning book~\cite{DL-Book}), where the algorithm memorizes the training samples without generalizing to the test set. The symptoms of overfitting can be alleviated by using regularization techniques, such as dropout and by limiting the number of training iterations (early stopping). However, in practice, an extremely large labeled dataset (typically a million samples) is required to train these randomly initialized deep CNNs for optimal performance.
	
	The \texttt{Gravity Spy} crowd-sourcing project mobilizes citizen scientists to hand-label spectrograms obtained from aLIGO time-series data. The rationale for this approach is that humans are capable of distinguishing different classes of noise from spectrograms after being shown only a few examples, which indicates that generic pattern recognizers developed in humans for real-world object recognition are also useful when distinguishing spectrograms of glitches. Therefore, this motivated us to apply a similar approach where a CNN is first trained on a large database of labeled real-world objects, and this knowledge is subsequently transferred by re-training (fine-tuning) the CNN on the dataset of spectrograms. In machine learning literature, this idea is referred to as transfer learning, which we describe in the following Section.

	\subsection{Transfer Learning}
	
	Transfer learning is an essential ingredient for artificial intelligence, where knowledge learned in one domain for some task can be transferred to another domain for a different task~\cite{Transferable}. In the context of deep learning, transfer learning is commonly implemented by pre-training a deep neural network, e.g., a CNN, on a large labeled dataset followed by fine-tuning (continued training) on a different dataset of interest, which is usually smaller. This approach has been successfully applied in many areas of computer vision~\cite{CNNFeatures}.
	
	It is well known that the initial layers of a CNN always learn to extract simple generic features (e.g., edges, corners, curves, color blobs, etc.), which are applicable to all types of images, whereas the final layers represent highly abstract and data-specific features~\cite{CNNFeatures}. Therefore, using a model optimally-trained on a large dataset and then fine-tuning it on a different dataset is expected to result in higher accuracy and a faster training process, compared to training the same CNNs from scratch, due to the shared features present in the initial layers.
	
	To demonstrate the power of transfer learning for classifying glitches, we compare the performance of the most popular CNN models for object recognition, namely Inception~\cite{arxiv:Sz} version 2 and 3, ResNet~\cite{arxiv:Km}, and VGG~\cite{arxiv:karen}, all of which were leading entries in recent ILSVRC competitions. These CNNs were optimally trained on a large dataset of images --- i.e., \texttt{ImageNet}~\cite{cvpr:Deng}, which contains 1.2 million labeled images of real-world objects belonging to 1000 categories (see~\Fref{fig:image} for sample images) --- over the course of 2 to 3 weeks using multiple GPUs by other research groups. We obtained the open-source weights from these models, and used them to initialize the CNNs, before fine-tuning each model on the dataset of glitches. 
	
	We show that transfer learning allows us to apply these powerful CNNs for classifying glitches using a very small training dataset of spectrograms to obtain significantly higher accuracies, reduce the training time by several orders of magnitude, and eliminate the need for optimizing hyper-parameters. Furthermore, we show how information from multiple duration spectrograms can be efficiently encoded into a single image in different color channels to enhance information provided to these CNN models.

	\subsection{Dataset}
	
\begin{figure*}
		\centering     
		{\includegraphics[width=1.07\textwidth]{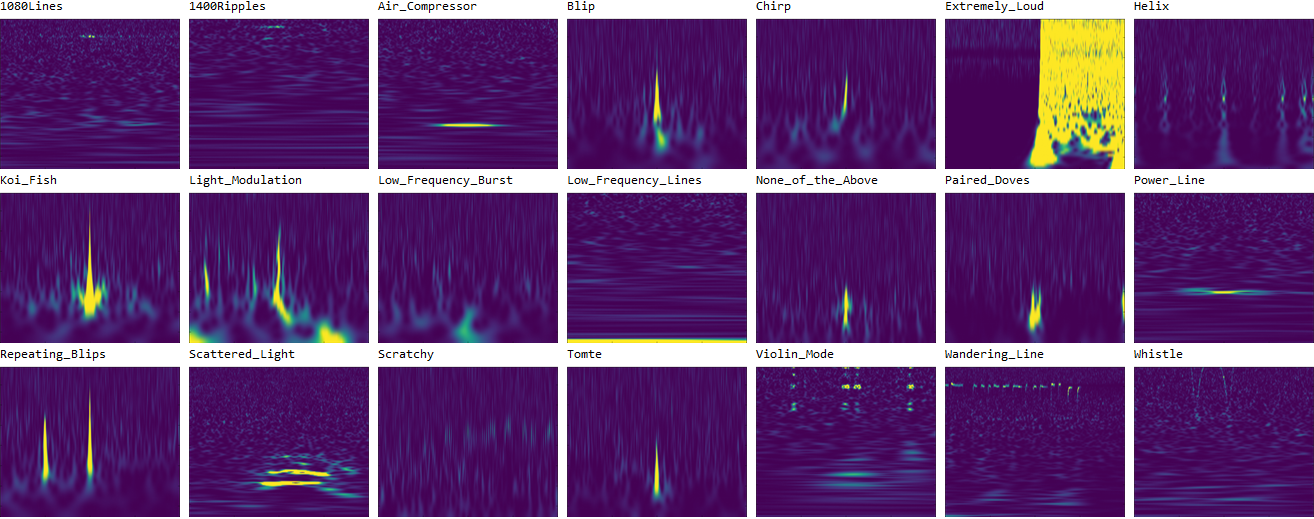}}		
		\caption{Classes of glitches in the \texttt{Gravity Spy} dataset from the first observing run of aLIGO. The \texttt{No\_Glitch} class is omitted in the figure. For each glitch sample, spectrograms with durations of 0.5s, 1.0s, 2.0s, and 4.0s are available. The objective is to predict the classes given the images.}
		\label{fig:glitches}
	\end{figure*}

	The \texttt{Gravity Spy} dataset, from the first observing run of LIGO, contains labeled spectrogram samples from 22 classes of glitches shown in~\Fref{fig:glitches}. Within the dataset, each sample contains 4 images recorded with durations: 0.5s, 1.0s, 2.0s, and 4.0s. These images were hand-labeled by citizen scientists participating in the \texttt{Gravity Spy} project, and the accuracy of the labeling was greatly enhanced by cross-validation techniques within the \texttt{Gravity Spy} infrastructure.
	
	We cropped each image closely around its frame, removing axis labels and white-spaces. Subsequently, the resolution of these images were $481\times575\times3$, with the third dimension corresponding to the RGB color channels. For training and testing, these images were down-sampled to the optimal resolution for each CNN model, either $224\times224\times3$ or $299\times299\times3$. We used the recent version of the  \texttt{Gravity Spy} dataset as of April 19, 2017, which has been slightly modified since the previous publications~\cite{spy:2016arXiv,multi:2017arXiv}, with two additional classes, namely 1080\_Lines and 1400\_Ripples, and about 300 extra elements in the Violin\_Mode class. We randomly split this dataset, containing about 8500 elements, into two parts such that approximately 80\% of samples in each class was in the training set, and 20\% of each class was in the testing set. 
	
	We did not use a separate validation set, since we did not engage in any kind of hyperparameter optimization that could have led to overfitting on the test set. This allowed us to use a larger fraction of the data for training and testing. We directly used the same hyperparameters as the pre-trained models.  For cross-verification of the performance, the Inception models were trained independently from the rest of the models, using a different deep learning software framework and different randomized test-train splits of the data.
	
	For training VGG16, VGG19, and ResNet50, we only used spectrogram images with 2.0s duration, since they contained sufficient information to resolve most of the long and short duration glitches. Since the dataset is small, image augmentation was used to artificially enhance the size of the dataset (except when training the Inception models). For each input image, we subtracted the mean of the training samples, then randomly shifted it by 0 to 10 pixels in horizontal and vertical directions, and randomly zoomed in and out with factors up to 0.05$\times$. The mean image of all the training samples was also subtracted from each sample used for testing. 
	
	For the Inception models, we found that we could improve the accuracy, without any image augmentation, by encoding multiple-duration spectrograms of each glitch into each of the RGB color channels to present maximum information to the CNNs. The 1.0s, 2.0s, and 4.0s duration spectrograms were converted to gray-scale, and then merged to produce RGB images with each channel encoding the different durations. Examples of these encoded images are shown in~\Fref{fig:color}. While previous work had used concatenated or parallel-views of multiple-duration spectrograms as inputs to specially designed CNNs trained from scratch~\cite{multi:2017arXiv}, our encoding method has the advantage of allowing the direct use of pre-trained CNN architectures designed for ImageNet for classification of multiple-duration spectrograms.

	\subsection{Training}
	
	The final fully-connected layer in each CNN model was replaced with another fully-connected layer having 22 neurons corresponding to each glitch class. To reduce overfitting, we added several dropout layers~\cite{Dropout} (which randomly turn neurons off to prevent co-adaptation between neurons and thus improves their ability to generalize) at the beginning of the network and right before the output layers. Apart from this, we also added one additional layer after the first trainable convolutional layer for ResNet. The details about network structures are listed in~\Tref{nets} of~\ref{apna}. The softmax function is used as the final layer in each model to provide probabilities of each class as the outputs.
	
	When training each model, we utilized the standard cross entropy loss (cost) function due to its good performance in classification problems in combination with the softmax layer~\cite{DL-Book}. We fine-tuned across all the layers since the dataset of glitches is very different from the objects in the ImageNet data. Each model was trained up to 100 epochs with the usual iterative procedure~\cite{DL-Book} and check-pointed after each epoch. The best performing check-points were chosen subsequently.
	
	We used the standard AdaGrad~\cite{op:adagrad} and ADAM~\cite{ADAM} optimizers as the learning algorithm~\cite{DL-Book} for training. Specifically, we used AdaGrad for ResNet, VGG16 and VGG19 with initial learning rate set to $10^{-4}$ and $\epsilon = 10^{-6}$ and the ADAM method with parameters $\beta_1=0.9$ and $\beta_2=0.999$ for the Inception models. Inception V2 and V3 models were implemented in the Wolfram Language~\cite{wolfram}, which internally uses the open-source MXNet~\cite{MXNet} framework for training. VGG and ResNet models were implemented and trained independently on a different randomized train-test split using TensorFlow~\cite{TensorFlow} via Keras~\cite{chollet2015keras}.
	
	\section{Results}
	\label{exp}
	
	A naive implementation of a logistic regression classifier on top of features extracted from the spectrograms using Inception models pre-trained on ImageNet, produced over 95\% accuracy without fine-tuning any layers. This strongly indicated that features learned for real-world object recognition are also directly useful for classification of spectrograms of LIGO noise transients, and thus offered incentive to now train (fine-tune) every layer. The results after fine-tuning are presented below.
	
	For the CNNs implemented in this paper, both InceptionV2 and InceptionV3 achieved over 98\% accuracy in fewer than 10 epochs of training (less than 20 minutes), VGG16 and VGG19 achieved over 98\% accuracy within 30 epochs of training. However, ResNet50 took much longer to converge. We compare the results of these CNNs trained with the transfer learning method with the CNN models trained from scratch~\cite{spy:2016arXiv,multi:2017arXiv} in Table~\ref{results}.  We found that each of our models consistently achieved over 98\% accuracy for many epochs, thus indicating that the performance is robust, regardless of the stopping criteria, and therefore the model is not overfitting on the test set.
	
	The precision and recall obtained with each model on every class is reported in~\ref{apna}. With InceptionV3, we achieved \textit{perfect} precision and recall on 8 classes: \texttt{1080Lines}, \texttt{1400Ripples}, \texttt{Air\_Compressor}, \texttt{Chirp}, \texttt{Helix}, \texttt{Paired\_Doves}, \texttt{Power\_Line}, and \texttt{Scratchy}.  With ResNet50, we achieved perfect precision and recall on 7 classes: \texttt{1080Lines}, \texttt{1400Ripples}, \texttt{Extremely\_Loud}, \texttt{Helix}, \texttt{Paired\_Doves}, \texttt{Scratchy}, and \texttt{Violin\_Mode}.
	
\begin{figure*}
		\centering     
		{\label{fig:e12}\includegraphics[width=160mm]{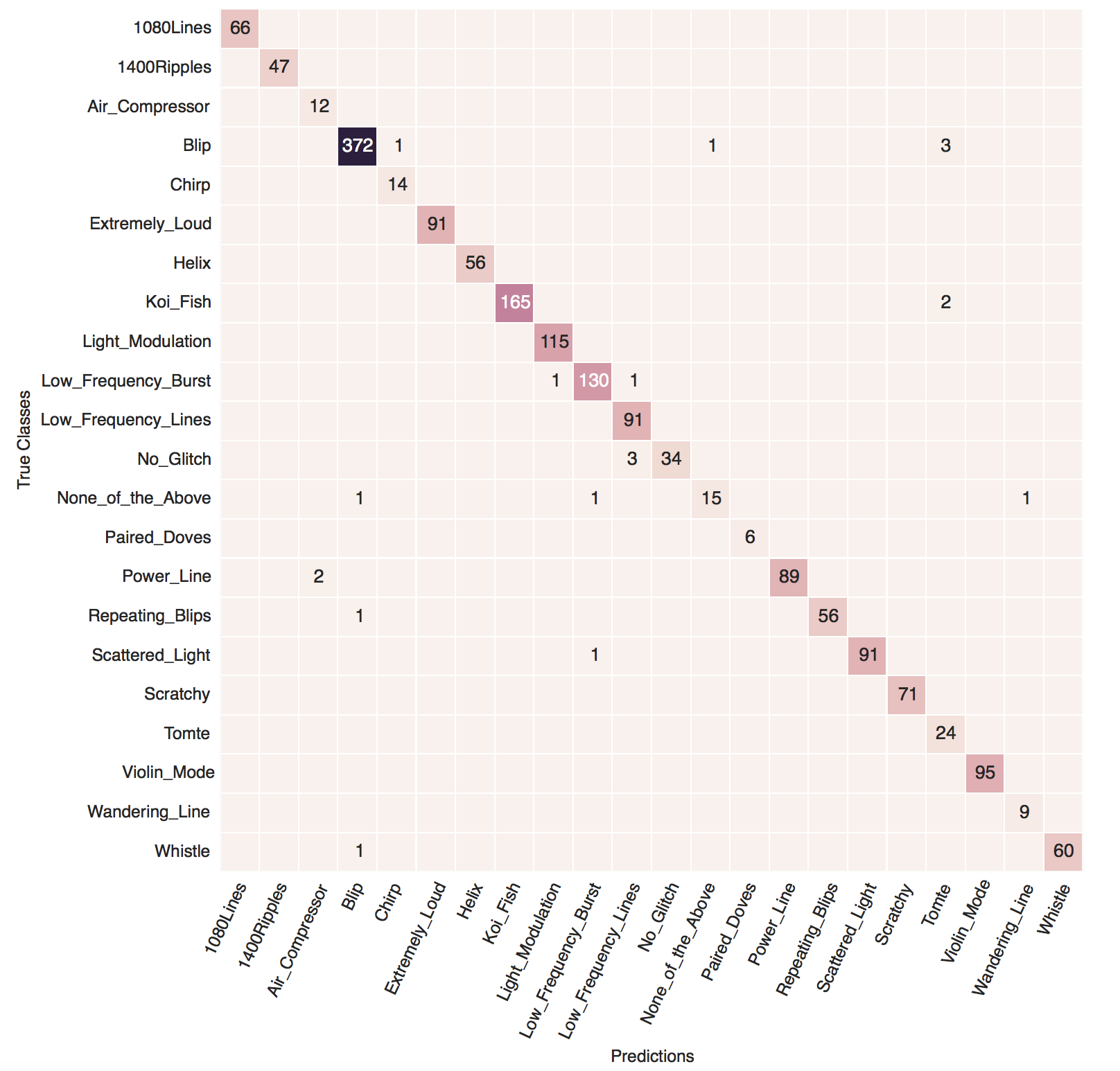}}	
		\begin{minipage}{0.65\textwidth} 
			
		\end{minipage}	
		\caption{Confusion matrix of ResNet50 after fine-tuning on the dataset of glitches. The accuracy is 98.84\%. Note that ResNet is the current state-of-the-art CNN model for a variety of image processing tasks including object recognition.}
		\label{fig:confusion}
	\end{figure*}

\begin{figure*}
		\centering     
		\includegraphics[width=163mm]{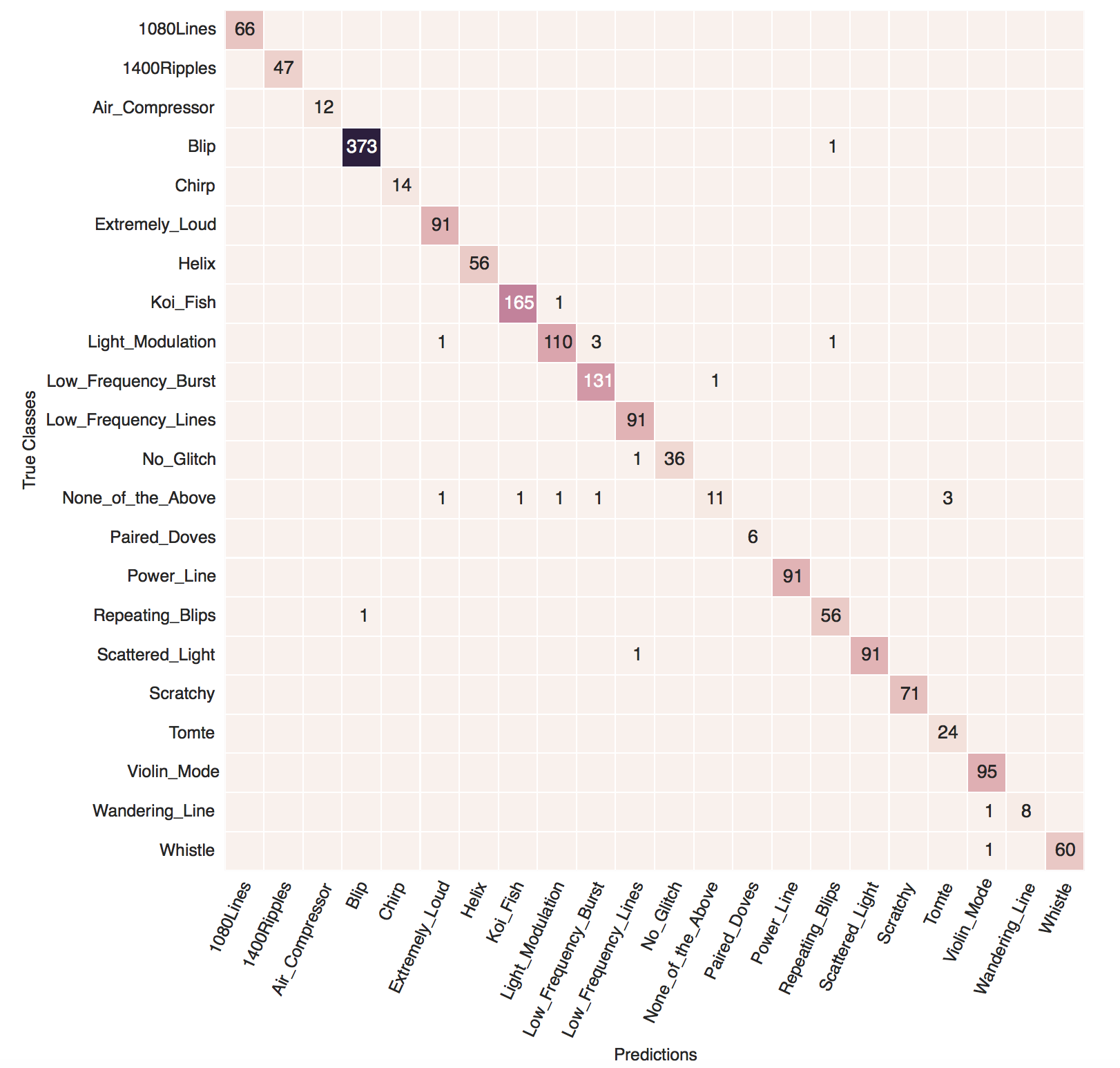}
		\begin{minipage}{0.65\textwidth} 
			
		\end{minipage}
		\caption{Confusion matrix for InceptionV3. The accuracy is also 98.84\%. It can be seen that different errors were made compared to ResNet50. Note that the \texttt{Chirp} glitch class (which is also the shape for true GW signals from mergers of binaries) and the \texttt{Paired\_Doves} class (which only had 24 training examples) was identified with perfect precision and recall.}
		\label{fig:confusionGoogv3}
	\end{figure*}

	Both ResNet50 and InceptionV3 achieved the highest accuracy of 98.84\% on the test set despite being trained independently via different methods on different splits of the data with 2.0s scans and RGB encoded scans respectively. They also obtained 100\% accuracy when considering the top-5 predictions, which implies that given any input, the true class can be narrowed down to within 5 classes with 100\% confidence. This is particularly useful, since the true class of a glitch is often ambiguous to even human experts. We have shown the confusion matrices for ResNet50 and InceptionV3 in~\Fref{fig:confusion} and~\Fref{fig:confusionGoogv3}, respectively. Confusion matrices of the remaining models are included in~\ref{apna}. The accuracies of each of these models, and previously published CNN models trained from scratch on the same dataset are reported in ~\Tref{results}. Note that each of our models consistently under-performed when trained with random initializations, i.e., without applying transfer learning. Therefore, it is evident that the networks trained with the transfer learning approach achieve better results than the CNN models trained from the ground up on the same dataset.
	
	
	\begin{table}
		
		\caption{\label{results}Accuracy on the Test Set}
		\footnotesize
		\begin{center}
			\begin{tabular}{@{}c c c c c c c c} 
				\br
				Neural Network & Top-1(20 classes) & Top-1 & Top-2 & Top-3 & Top-4 & Top-5 \\  
				\mr
				
				Tuned-VGG16 & 98.02\% & 98.15\% & 99.36\% & 99.71\% & 99.83\% & 99.88\% \\
				Tuned-VGG19 & 98.08\% & 98.21\% & 99.31\% & 99.60\% & 99.71\% & 99.71\%  \\
				Tuned-ResNet50& 98.76\% & 98.84\% & 99.71\% & 99.83\% & 99.94\% & 100\%  \\
				Tuned-InceptionV2& 98.70\% & 98.78\% & 99.59\% & 99.71\% &  99.94\% & 100\%  \\
				Tuned-InceptionV3& 98.76\% & 98.84\% & 99.71\% & 99.88\% & 99.94\% & 100\%  \\
				CNN in~\cite{spy:2016arXiv,multi:2017arXiv} &96.72\%& 96.70\%&98.32\%&99.13\%&99.31\%&99.36\% \\
				\br 
			\end{tabular}
		\end{center}
		The table lists the top-1 to top-5 accuracies for different CNNs on the testing set. The top-1 accuracy for 20 classes (excluding \texttt{1080Lines} and \texttt{1400Ripples}) is also shown for comparison with previous publications. We re-trained the 4 layer merged-view CNN model described in the publications~\cite{spy:2016arXiv,multi:2017arXiv} from scratch on our same train-test dataset for a fair comparison, since the Gravity Spy dataset has been recently updated. Note that our Inception and ResNet models are capable of narrowing down any input to within 5 classes with 100\% accuracy.
	\end{table}
	\normalsize

	\section{Discussion}
	\label{Dis}
	
	The results presented in~\Tref{results} indicate that transfer learning with state-of-the-art image classification CNNs achieves optimal performance, while significantly reducing the training time to less than an hour, and eliminating the difficulties associated with hyperparameter tuning. Employing CNN models with very deep and complex structures is feasible with this approach on very small datasets, since the larger training set used for pre-training enables learning good feature extractors in the initial layers, and thus prevents overfitting before learning good features. In addition, the CNNs were able to learn high level features specific to the small dataset of glitches during the fine-tuning process throughout all layers, which further improved the classification results.
	
	We have also provided a comparison between the most popular state-of-the-art models for image classification on the transfer learning problem. Furthermore, we have demonstrated that transfer learning always provides significant improvements over the traditional approach used for classifying spectrograms, and allows the use of the most powerful CNNs on very small datasets, even when they are very dissimilar compared to the original data used for pre-training. Once trained, inference (evaluation on new inputs) can be carried out in a few milliseconds using a GPU, allowing real-time classification of glitches in aLIGO.

	Note that both the ResNet and Inception models achieved perfect precision and recall on the \texttt{Paired\_Doves} test set even though it was the smallest class with only ~21 elements in the training set. Previous methods trained from the ground up had achieved sub-optimal results for this class~\cite{spy:2016arXiv}. This analysis strongly suggests that the high accuracy is due to transfer learning from the larger dataset, since similar patterns could have been learned when pre-trained on the ImageNet data. Therefore, we have shown that, although the number of samples vary greatly between different classes (up to roughly a factor of 100), we could obtain high precision and recall on even the smallest classes without employing any class balancing techniques for training.
	
	The best performing models on ImageNet are currently based on ensembles/committees of different CNNs. Since the confusion matrices indicate that each CNN has different strengths and weaknesses for many classes, we expect that using an ensemble of different models would further boost the accuracy for glitch classification. Furthermore, as the Gravity Spy dataset is enlarged in the future with more hand-labeled glitches, our CNNs may be re-trained with the largest available dataset to improve the accuracy. 
	
	However, there may be an upper bound set by the error-rate of humans providing the labels. It can be seen in~\Fref{fig:misclassified} that most of the misclassification made by our CNNs were either due to incorrect labeling, or due to superposition of two different classes of glitches, or due to inputs whose true class remain ambiguous. Since our method can be trained very quickly, it can be used to efficiently correct mislabeled elements in the original dataset by training on a large number of randomized train-test splits, and searching for the most commonly mislabeled elements among all these test sets.
	
	The CNNs we have developed in this study may be immediately used in LIGO pipelines for classifying known categories of glitches with high accuracy in real-time. The well-known ability of deep neural networks to generalize implies that the classifications would be resilient to changes in background noise. Therefore, we expect the excellent performance we achieved on the current data would translate to future observing runs. Nonetheless, the accuracy can be further improved by periodically re-training with larger datasets containing all available labeled examples of glitches at each time.
	
	After applying transfer learning, i.e., fine-tuning the CNNs on the small dataset of glitches, the CNNs may also be used as good feature extractors for finding new categories of glitches from unlabeled data in an unsupervised or semi-supervised manner. This method, which combines supervised learning, transfer learning, and unsupervised learning, can be used to identify many more categories of noise transients and estimate at what times new types of glitches with similar morphologies start occurring. This may also be used to correct mislabeled glitches in the original dataset used for training/testing by searching for anomalies in the feature-space.
	
	For instance, consider our InceptionV3 model: removing the final softmax and fully-connected layer near the output produces a CNN that maps any input image to a 1008-dimensional vector. In this high-dimensional space, glitches having similar morphology will be clustered together. Anomalies and mis-labeled examples will appear isolated from the clusters. This can be visualized by reducing the dimensions of the feature space to 2-D or 3-D using a suitable dimension reduction algorithm (e.g., t-SNE~\cite{TSNE}) as shown in~\Fref{fig:tSNE1} and~\Fref{fig:tSNE2}). 
	
	Therefore, when new types of glitches appear, which are classified as \texttt{None\_of\_the\_Above} by our CNN model, they may be mapped to vectors using these truncated CNN feature extractors. Then, new classes of glitches may be identified by applying standard unsupervised clustering algorithms (see Appendix C for a proof-of-concept). This automated clustering can also be used to accelerate human labeling of new glitches in a semi-supervised manner by presenting batches of similar glitches to a citizen scientist.
	
	We believe the reason why our clustering technique works very well is because we have used very deep CNNs that are capable of extracting and representing a large number of highly abstract features of the input images. Therefore, transfer learning enables the use of these CNNs which are excellent feature extractors, thus effectively enabling the use of unsupervised learning techniques for finding new classes of glitches.  The performance of different clustering methods may be investigated in detail in a subsequent study.
	
	\section{Conclusion}
	\label{end}
	
	This article is part of a program that aims to spread the use of deep learning in many related areas of multimessenger astrophysics~\cite{dnn:2017a}, ranging from detection and characterization of GW sources, to detector characterization. This is the first application of Deep Transfer Learning in the context of aLIGO. Using this method, we have developed a state-of-the-art DNNs for aLIGO glitch classification using the \texttt{Gravity Spy} dataset. We have shown that by combining the pre-trained weights of state-of-the-art CNNs with fine-tuning across all layers, we can further increase the accuracy compared to neural networks that have been trained from scratch using only the \texttt{Gravity Spy} dataset. Furthermore, we have shown that the training time is significantly reduced with our approach when compared to traditional methods and the effort required to design CNN models and optimize hyper-parameters can be eliminated.
		
	Note that even though the \texttt{Gravity Spy} dataset of spectrograms was not at all similar to the real-world objects found in the ImageNet database, we have demonstrated that the transfer learning approach works extremely well. Therefore, we expect that our method will be useful for time-series classification in general, where the problem can be represented in the form of images. Furthermore, we have developed a novel method to encode different duration spectrograms (i.e., multiple views of the same input) as the different RGB color channels in a single image, which renders excellent results and allows the direct use of pre-trained state-of-the-art CNN models.
	
	The algorithms we have introduced in this article may be used to classify new time-series data in the \texttt{Gravity Spy} project, as well as data from future aLIGO observing runs and data from international partners, such as Virgo~\cite{Virgo:2015}, KAGRA~\cite{Hiroshe:2014} and LIGO-India~\cite{Unni:2013} as they come online in the next few years. This would be useful both for improving the data quality the detectors and for generating vetoes in GW search pipelines. The transfer learning method allows us to use very deep CNNs, with well-known architectures, which are capable of extracting more complex and abstract features from the spectrogram images. New classes of glitches may also be found in an unsupervised manner or labeled rapidly in a semi-supervised manner with our method by truncating these CNNs, and using them as feature extractors for clustering algorithms. 
	
	By employing this transfer learning method for glitch classification, we can create a larger dataset of labeled transients with real aLIGO data, and the corresponding time-series can be fetched from the GPS times to produce labeled time-series inputs. This dataset may be added to the training process in the Deep Filtering pipeline introduced in~\cite{dnn:2017a}, to unify detection, classification, and parameter estimation directly from time-series data streams for GW detectors in \textit{real-time}. The fully-trained CNNs developed in this article will be made available to the community for use in glitch classification and detector characterization pipelines.
	
	\section{Acknowledgements}
	\label{ack}
	We are grateful to NVIDIA for supporting this research with GPU hardware donations, to Wolfram Research for offering several Mathematica licenses, and to Vlad Kindratenko for providing dedicated access to a machine at the Innovative Systems Lab at NCSA. We acknowledge the Gravity Spy project and the citizen scientists who participated in it for creating the dataset we used. We thank Gabrielle Allen, Scott Coughlin, Vicky Kalogera, Joshua Smith, Kai Staats, Sara Bahaadini, and Michael Zevin for productive interactions. We thank Kai Staats, Laura Nuttall, and the Gravity Spy team for reviewing this article and providing feedback. We also acknowledge the LIGO collaboration for the use of computational resources and for conducting the experiments from which the raw data was obtained. This document has LIGO number P1700128.

	\vspace{2pc}
	
	\bibliographystyle{iopart-num}
	\bibliography{references}
	
	\clearpage
	
	\appendix
	\section{}
	\label{apna}
	
\begin{figure*}[htp!]
		\centering     
		\includegraphics[width=160mm]{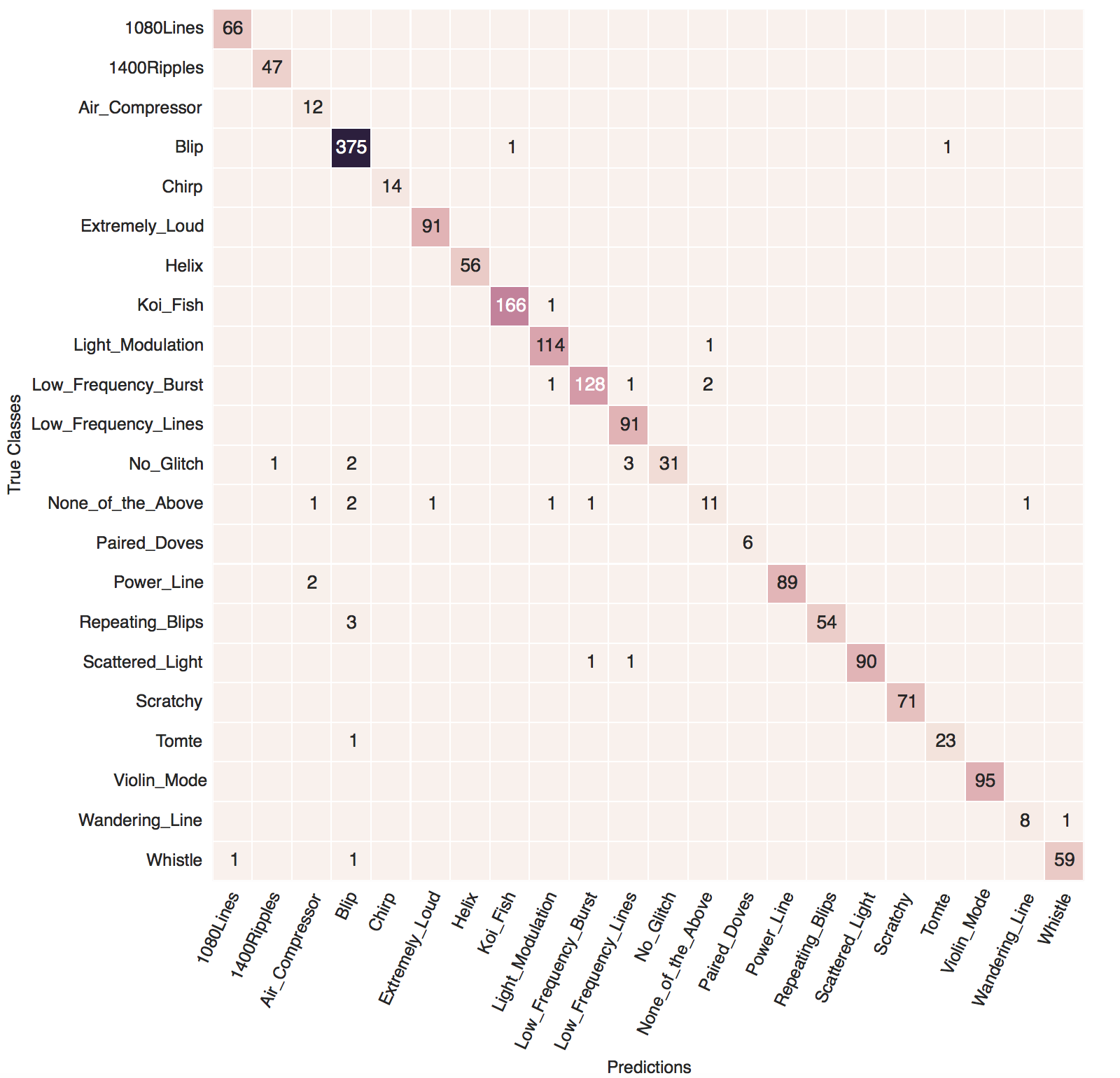}
		\begin{minipage}{0.65\textwidth} 
		\end{minipage}
		\caption{Confusion matrix for VGG16. The accuracy is 98.15\%.}
		\label{fig:confusionVGG16}
	\end{figure*}

\begin{figure}
		\centering     
		\includegraphics[width=160mm]{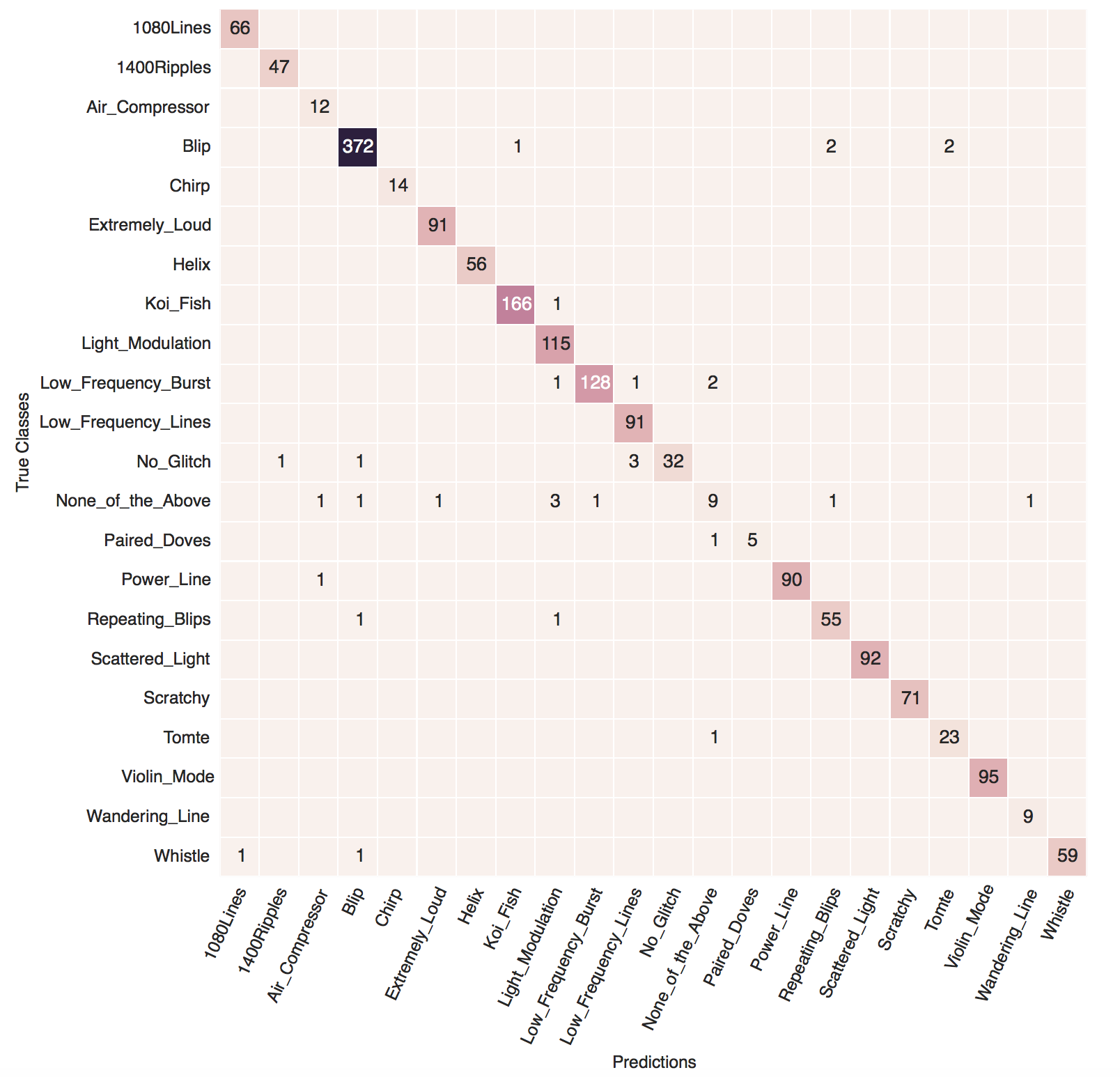}
		\begin{minipage}{0.65\textwidth} 
		\end{minipage}
		\caption{Confusion matrix for VGG19. The accuracy is 98.21\%.}
		\label{fig:confusionVGG19}
	\end{figure}

\begin{figure}[h]
		\centering     
		\includegraphics[width=160mm]{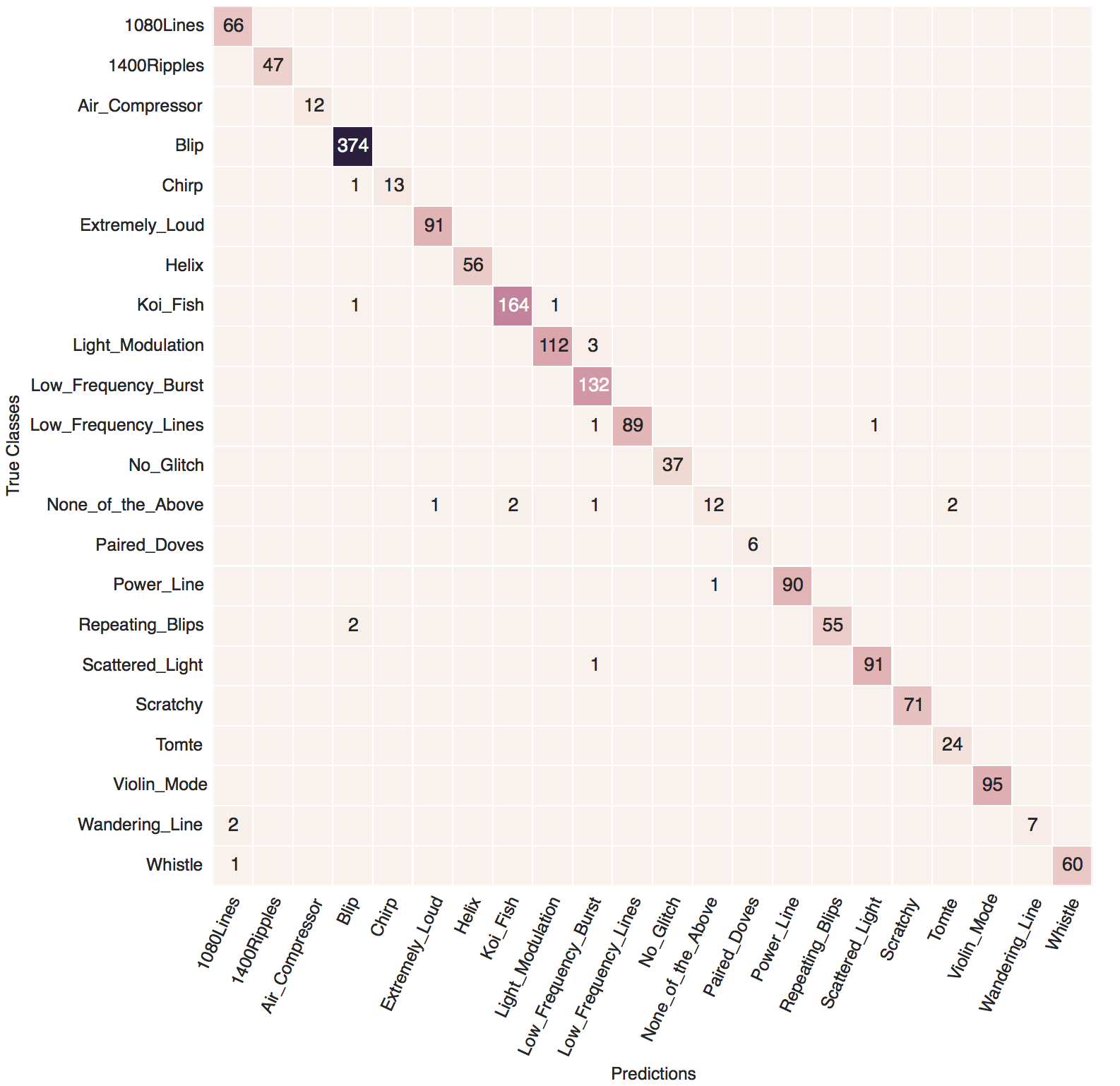}
		\begin{minipage}{0.65\textwidth} 
		\end{minipage}
		\caption{Confusion matrix for InceptionV2. The accuracy is 98.78\%.}
		\label{fig:confusionGoog}
	\end{figure}

	In this Section we present supplementary information summarizing the networks we experimented with to classify spectrograms of the \texttt{Gravity Spy} dataset. \Fref{fig:confusionVGG16}--~\Fref{fig:confusionGoog} present the confusion matrices for VGG16, VGG19 and InceptionV2 models respectively.

	\begin{figure*}
		\centering     
		{\includegraphics[width=1.0\textwidth]{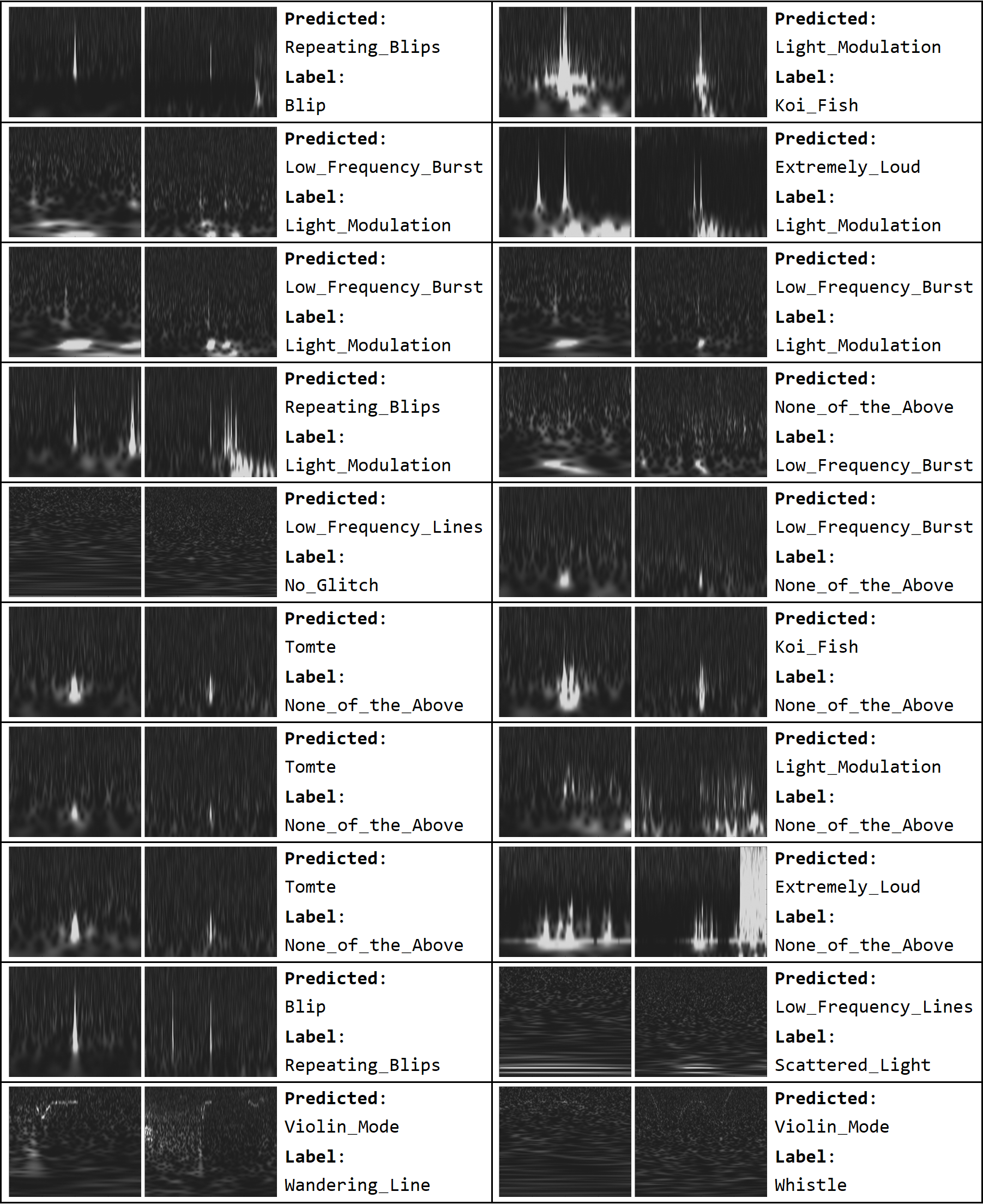}}		
		\caption{All misclassified elements in the test set for the InceptionV3 model. The 1s and 4s spectrograms are shown. It can be seen that the CNNs always outputs reasonable classifications with many of the mistakes due to incorrect labels in the test set or glitches whose classes appear ambiguous to even humans. Therefore, the CNN has achieved close to ideal accuracy, i.e, the Bayes error rate. This suggests that mislabeled examples can be found and corrected by trying different randomized test-train splits of the dataset, and then inspecting the mistakes made by the CNN on the test set.}
		\label{fig:misclassified}
	\end{figure*}

	\noindent \Tref{recall_precision} presents the recall and precision of our Deep Transfer Learning algorithm for glitch classification. These results clearly exhibit the power of this method, especially for rare classes with few labeled examples, therefore the transfer learning method offers a promising framework for future development in glitch classification for GW detectors.
	\begin{table*}[htp!]
		\caption{\label{recall_precision}Table of Precision and Recall on the Test Set}
		\scriptsize
		\begin{center}
			\begin{tabular}{@{}c c c c c c c c} 
				\br
				Class & Quantity  & VGG16 & VGG19 & ResNet50 & InceptionV2  & InceptionV3  &  CNN in~\cite{spy:2016arXiv,multi:2017arXiv} \\  
				\mr
				1080Lines& $\begin{tabular}{c}recall\\ \hline precision \\ \end{tabular}$ & $\begin{tabular}{c}100\%\\ \hline 98.51\% \\ \end{tabular}$ &$\begin{tabular}{c}100\%\\ \hline 98.50\% \\ \end{tabular}$& $\begin{tabular}{c}100\%\\ \hline 100\% \\ \end{tabular}$ & $\begin{tabular}{c}100\%\\ \hline 95.65\% \\ \end{tabular}$ &$\begin{tabular}{c}100\%\\ \hline 100\% \\ \end{tabular}$ & $\begin{tabular}{c}100\%\\ \hline 97.06\% \\ \end{tabular}$ \\ \hline
				
				1400Ripples& $\begin{tabular}{c}recall\\ \hline precision \\ \end{tabular}$ & $\begin{tabular}{c}100\%\\ \hline 97.92\% \\ \end{tabular}$& $\begin{tabular}{c}100\%\\ \hline 97.91\% \\ \end{tabular}$ & $\begin{tabular}{c}100\%\\ \hline 100\% \\ \end{tabular}$ & $\begin{tabular}{c}100\%\\ \hline 100\% \\ \end{tabular}$ &$\begin{tabular}{c}100\%\\ \hline 100\% \\ \end{tabular}$ & $\begin{tabular}{c}91.49\%\\ \hline 82.69\% \\ \end{tabular}$ \\ \hline
				
				Air\_Compressor& $\begin{tabular}{c}recall\\ \hline precision \\ \end{tabular}$ & $\begin{tabular}{c}100\%\\ \hline 80.00\% \\ \end{tabular}$ & $\begin{tabular}{c}100\%\\ \hline 85.71\% \\ \end{tabular}$ & $\begin{tabular}{c}100\%\\ \hline 85.71\% \\ \end{tabular}$  & $\begin{tabular}{c}100\%\\ \hline 100\% \\ \end{tabular}$ &$\begin{tabular}{c}100\%\\ \hline 100\% \\ \end{tabular}$ & $\begin{tabular}{c}100\%\\ \hline 85.71\% \\ \end{tabular}$\\ \hline
				
				Blip& $\begin{tabular}{c}recall\\ \hline precision \\ \end{tabular}$  &$\begin{tabular}{c}99.47\%\\ \hline 97.66\% \\ \end{tabular}$ &$\begin{tabular}{c}98.67\%\\ \hline 98.94\% \\ \end{tabular}$ & $\begin{tabular}{c}98.67\%\\ \hline 99.20\% \\ \end{tabular}$ & $\begin{tabular}{c}99.99\%\\ \hline 98.94\% \\ \end{tabular}$  &$\begin{tabular}{c}99.73\%\\ \hline 99.73\% \\ \end{tabular}$ & $\begin{tabular}{c}98.94\%\\ \hline 98.15\% \\ \end{tabular}$  \\ \hline
				
				Chirp&  $\begin{tabular}{c}recall\\ \hline precision \\ \end{tabular}$ & $\begin{tabular}{c}100\%\\ \hline 100\% \\ \end{tabular}$& $\begin{tabular}{c}100\%\\ \hline 100\% \\ \end{tabular}$  & $\begin{tabular}{c}100\%\\ \hline 93.33\% \\ \end{tabular}$ &$\begin{tabular}{c}92.86\%\\ \hline 100\% \\ \end{tabular}$&$\begin{tabular}{c}100\%\\ \hline 100\% \\ \end{tabular}$ & $\begin{tabular}{c}100\%\\ \hline 100\% \\ \end{tabular}$\\ \hline
				
				Extremely\_Loud& $\begin{tabular}{c}recall\\ \hline precision \\ \end{tabular}$  & $\begin{tabular}{c}100\%\\ \hline 98.91\% \\ \end{tabular}$ & $\begin{tabular}{c}100\%\\ \hline 98.91\% \\ \end{tabular}$ & $\begin{tabular}{c}100\%\\ \hline 100\% \\ \end{tabular}$  &$\begin{tabular}{c}100\%\\ \hline 98.91\% \\ \end{tabular}$ &$\begin{tabular}{c}100\%\\ \hline 97.85\% \\ \end{tabular}$ & $\begin{tabular}{c}100\%\\ \hline 100\% \\ \end{tabular}$\\ \hline
				
				Helix& $\begin{tabular}{c}recall\\ \hline precision \\ \end{tabular}$ & $\begin{tabular}{c}100\%\\ \hline 100\% \\ \end{tabular}$& $\begin{tabular}{c}100\%\\ \hline 100\% \\ \end{tabular}$  & $\begin{tabular}{c}100\%\\ \hline 100\% \\ \end{tabular}$ &$\begin{tabular}{c}100\%\\ \hline 100\% \\ \end{tabular}$&$\begin{tabular}{c}100\%\\ \hline 100\% \\ \end{tabular}$ & $\begin{tabular}{c}98.21\%\\ \hline 94.83\% \\ \end{tabular}$\\ \hline
				
				Koi\_Fish& $\begin{tabular}{c}recall\\ \hline precision \\ \end{tabular}$ & $\begin{tabular}{c}99.40\%\\ \hline 99.41\% \\ \end{tabular}$& $\begin{tabular}{c}99.40\%\\ \hline 99.40\% \\ \end{tabular}$ & $\begin{tabular}{c}98.80\%\\ \hline 100\% \\ \end{tabular}$ &$\begin{tabular}{c}98.80\%\\ \hline 98.80\% \\ \end{tabular}$&$\begin{tabular}{c}99.40\%\\ \hline 99.40\% \\ \end{tabular}$ & $\begin{tabular}{c}100\%\\ \hline 99.40\% \\ \end{tabular}$ \\ \hline
				
				Light\_Modulation& $\begin{tabular}{c}recall\\ \hline precision \\ \end{tabular}$ & $\begin{tabular}{c}99.13\%\\ \hline 97.44\% \\ \end{tabular}$& $\begin{tabular}{c}100\%\\ \hline 95.04\% \\ \end{tabular}$  & $\begin{tabular}{c}100\%\\ \hline 99.14\% \\ \end{tabular}$ &$\begin{tabular}{c}97.39\%\\ \hline 99.11\% \\ \end{tabular}$&$\begin{tabular}{c}95.65\%\\ \hline 98.21\% \\ \end{tabular}$ & $\begin{tabular}{c}99.13\%\\ \hline 98.28\% \\ \end{tabular}$\\ \hline
				
				Low\_Frequency\_Burst& $\begin{tabular}{c}recall\\ \hline precision \\ \end{tabular}$ & $\begin{tabular}{c}96.97\%\\ \hline 98.46\% \\ \end{tabular}$& $\begin{tabular}{c}96.97\%\\ \hline 99.22\% \\ \end{tabular}$ & $\begin{tabular}{c}98.48\%\\ \hline 98.48\% \\ \end{tabular}$ &$\begin{tabular}{c}100\%\\ \hline 95.65\% \\ \end{tabular}$&$\begin{tabular}{c}99.24\%\\ \hline 97.04\% \\ \end{tabular}$ &$\begin{tabular}{c}96.21\%\\ \hline 100\% \\ \end{tabular}$  \\ \hline
				
				Low\_Frequency\_Lines& $\begin{tabular}{c}recall\\ \hline precision \\ \end{tabular}$ & $\begin{tabular}{c}100\%\\ \hline 97.49\% \\ \end{tabular}$& $\begin{tabular}{c}100\%\\ \hline 95.79\% \\ \end{tabular}$ & $\begin{tabular}{c}100\%\\ \hline 95.79\% \\ \end{tabular}$ &$\begin{tabular}{c}97.80\%\\ \hline 100\% \\ \end{tabular}$&$\begin{tabular}{c}100\%\\ \hline 97.85\% \\ \end{tabular}$ &$\begin{tabular}{c}95.60\%\\ \hline 92.55\% \\ \end{tabular}$ \\ \hline
				
				No\_Glitch& $\begin{tabular}{c}recall\\ \hline precision \\ \end{tabular}$ & $\begin{tabular}{c}83.78\%\\ \hline 100\% \\ \end{tabular}$& $\begin{tabular}{c}86.49\%\\ \hline 100\% \\ \end{tabular}$ & $\begin{tabular}{c}91.89\%\\ \hline 100\% \\ \end{tabular}$ &$\begin{tabular}{c}100\%\\ \hline 100\% \\ \end{tabular}$&$\begin{tabular}{c}97.30\%\\ \hline 100\% \\ \end{tabular}$ & $\begin{tabular}{c}86.49\%\\ \hline 94.12\% \\ \end{tabular}$ \\ \hline
				
				None\_of\_the\_Above& $\begin{tabular}{c}recall\\ \hline precision \\ \end{tabular}$ & $\begin{tabular}{c}61.11\%\\ \hline 78.57\% \\ \end{tabular}$& $\begin{tabular}{c}50.00\%\\ \hline 69.23\% \\ \end{tabular}$ & $\begin{tabular}{c}83.33\%\\ \hline93.75\% \\ \end{tabular}$ &$\begin{tabular}{c}66.67\%\\ \hline 92.31\% \\ \end{tabular}$&$\begin{tabular}{c}61.11\%\\ \hline 91.67\% \\ \end{tabular}$ & $\begin{tabular}{c}44.44\%\\ \hline 80.00\% \\ \end{tabular}$ \\ \hline
				
				Paired\_Doves& $\begin{tabular}{c}recall\\ \hline precision \\ \end{tabular}$ & $\begin{tabular}{c}100\%\\ \hline 100\% \\ \end{tabular}$& $\begin{tabular}{c}83.33\%\\ \hline 100\% \\ \end{tabular}$  & $\begin{tabular}{c}100\%\\ \hline 100\% \\ \end{tabular}$ &$\begin{tabular}{c}100\%\\ \hline 100\% \\ \end{tabular}$&$\begin{tabular}{c}100\%\\ \hline 100\% \\ \end{tabular}$ & $\begin{tabular}{c}100\%\\ \hline 100\% \\ \end{tabular}$\\ \hline
				
				Power\_Line& $\begin{tabular}{c}recall\\ \hline precision \\ \end{tabular}$ & $\begin{tabular}{c}97.80\%\\ \hline 100\% \\ \end{tabular}$& $\begin{tabular}{c}98.90\%\\ \hline 100\% \\ \end{tabular}$ & $\begin{tabular}{c}97.80\%\\ \hline 100\% \\ \end{tabular}$ &$\begin{tabular}{c}98.90\%\\ \hline 100\% \\ \end{tabular}$&$\begin{tabular}{c}100\%\\ \hline 100\% \\ \end{tabular}$ & $\begin{tabular}{c}98.90\%\\ \hline 100\% \\ \end{tabular}$ \\ \hline
				
				Repeating\_Blips&$\begin{tabular}{c}recall\\ \hline precision \\ \end{tabular}$ & $\begin{tabular}{c}94.74\%\\ \hline 100\% \\ \end{tabular}$& $\begin{tabular}{c}96.49\%\\ \hline 94.83\% \\ \end{tabular}$  & $\begin{tabular}{c}98.25\%\\ \hline 100\% \\ \end{tabular}$ &$\begin{tabular}{c}96.49\%\\ \hline 100\% \\ \end{tabular}$&$\begin{tabular}{c}98.25\%\\ \hline 96.55\% \\ \end{tabular}$ & $\begin{tabular}{c}89.47\%\\ \hline 94.44\% \\ \end{tabular}$ \\ \hline
				
				Scattered\_Light& $\begin{tabular}{c}recall\\ \hline precision \\ \end{tabular}$ & $\begin{tabular}{c}97.83\%\\ \hline 100\% \\ \end{tabular}$& $\begin{tabular}{c}100\%\\ \hline 100\% \\ \end{tabular}$  & $\begin{tabular}{c}98.91\%\\ \hline 100\% \\ \end{tabular}$ &$\begin{tabular}{c}98.91\%\\ \hline 98.91\% \\ \end{tabular}$&$\begin{tabular}{c}98.91\%\\ \hline 100\% \\ \end{tabular}$ & $\begin{tabular}{c}98.91\%\\ \hline 97.85\% \\ \end{tabular}$\\ \hline
				
				Scratchy&$\begin{tabular}{c}recall\\ \hline precision \\ \end{tabular}$ & $\begin{tabular}{c}100\%\\ \hline 100\% \\ \end{tabular}$& $\begin{tabular}{c}100\%\\ \hline 100\% \\ \end{tabular}$ & $\begin{tabular}{c}100\%\\ \hline 100\% \\ \end{tabular}$ &$\begin{tabular}{c}100\%\\ \hline 100\% \\ \end{tabular}$&$\begin{tabular}{c}100\%\\ \hline 100\% \\ \end{tabular}$ & $\begin{tabular}{c}100\%\\ \hline 97.26\% \\ \end{tabular}$  \\ \hline
				
				Tomte& $\begin{tabular}{c}recall\\ \hline precision \\ \end{tabular}$ & $\begin{tabular}{c}95.83\%\\ \hline 95.83\% \\ \end{tabular}$& $\begin{tabular}{c}95.83\%\\ \hline 92.00\% \\ \end{tabular}$  & $\begin{tabular}{c}100\%\\ \hline 82.76\% \\ \end{tabular}$ &$\begin{tabular}{c}100\%\\ \hline 92.31\% \\ \end{tabular}$&$\begin{tabular}{c}100\%\\ \hline 88.89\% \\ \end{tabular}$ & $\begin{tabular}{c}100\%\\ \hline 85.71\% \\ \end{tabular}$\\ \hline
				
				Violin\_Mode& $\begin{tabular}{c}recall\\ \hline precision \\ \end{tabular}$ & $\begin{tabular}{c}100\%\\ \hline 100\% \\ \end{tabular}$& $\begin{tabular}{c}100\%\\ \hline 100\% \\ \end{tabular}$ & $\begin{tabular}{c}100\%\\ \hline 100\% \\ \end{tabular}$ &$\begin{tabular}{c}100\%\\ \hline 100\% \\ \end{tabular}$&$\begin{tabular}{c}100\%\\ \hline 97.94\% \\ \end{tabular}$ & $\begin{tabular}{c}97.89\%\\ \hline 95.88\% \\ \end{tabular}$ \\ \hline
				
				Wandering\_Line& $\begin{tabular}{c}recall\\ \hline precision \\ \end{tabular}$ & $\begin{tabular}{c}88.89\%\\ \hline 88.89\% \\ \end{tabular}$& $\begin{tabular}{c}100\%\\ \hline 90.00\% \\ \end{tabular}$ & $\begin{tabular}{c}100\%\\ \hline 90.00\% \\ \end{tabular}$ &$\begin{tabular}{c}77.78\%\\ \hline 100\% \\ \end{tabular}$&$\begin{tabular}{c}88.89\%\\ \hline 100\% \\ \end{tabular}$ & $\begin{tabular}{c}66.67\%\\ \hline 75.00\% \\ \end{tabular}$ \\ \hline
				
				Whistle& $\begin{tabular}{c}recall\\ \hline precision \\ \end{tabular}$ & $\begin{tabular}{c}96.72\%\\ \hline 98.33\% \\ \end{tabular}$& $\begin{tabular}{c}96.72\%\\ \hline 100\% \\ \end{tabular}$  & $\begin{tabular}{c}98.36\%\\ \hline 100\% \\ \end{tabular}$ &$\begin{tabular}{c}98.36\%\\ \hline 100\% \\ \end{tabular}$&$\begin{tabular}{c}98.36\%\\ \hline 100\% \\ \end{tabular}$ & $\begin{tabular}{c}83.61\%\\ \hline 94.44\% \\ \end{tabular}$\\ 
				\br 
			\end{tabular}
		\end{center}
		\normalsize
		The table lists recall and precision for different CNN models. The last column compares the results with the merged-view CNN model described in~\cite{spy:2016arXiv,multi:2017arXiv} after training and testing from scratch on the same train-test split we used for our models with the updated Gravity Spy dataset.
	\end{table*}
	\clearpage
	
	\section{}
	
\begin{figure*}
		\centering     
		{\includegraphics[width=1.1\textwidth]{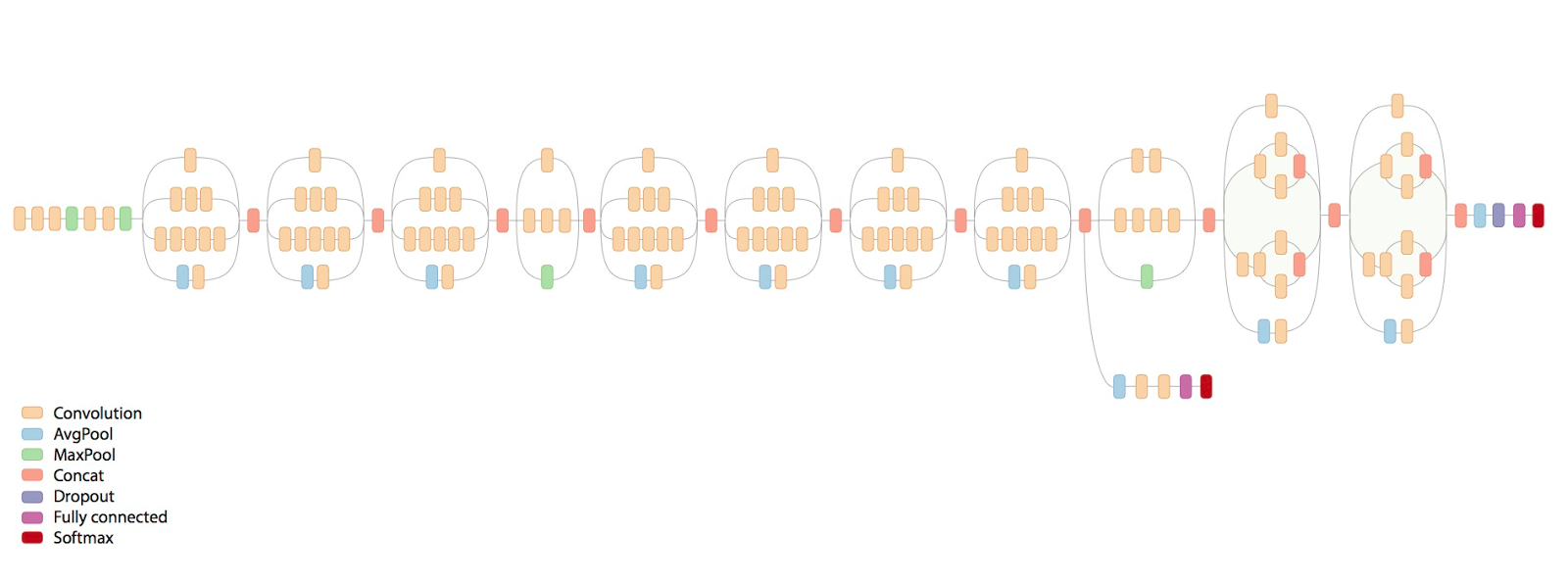}}		
		\caption{Architecture of InceptionV2 and InceptionV3 models. The parallel structure with Inception modules significantly improves evaluation speed and reduces total model size compared to the other CNN models. We used the InceptionV2 model currently in use for the ImageIdentify project pre-trained on a superset of the ImageNet dataset, which is publicly available at~\cite{ImageIdentify}.}
		\label{fig:Inc3}
	\end{figure*}

	\noindent \Tref{nets} describes in detail the architecture of VGG16, VGG19 and ResNet50, including a snapshot of the final structure of these networks. The more complex architecture of the Inception model is shown in~\Fref{fig:Inc3}.
	
	\begin{table}[h]
		
		\centering{\caption{ \label{nets} Neural Network Structures}}
		\footnotesize
		\begin{tabular}{@{}c c c } 
			\br
			VGG16 & VGG19 & ResNet50  \\
			\mr
			Conv3-64 $\times$ 2& Conv3-64 $\times$ 2 & $\left(
			\begin{tabular}{c}
			1 $\times$ 1, 64 \\
			3 $\times$ 3, 64 \\
			1 $\times$ 1, 256 
			\end{tabular}
			\right) \times 3$  \\ 
			\hline
			Conv3-128 $\times$ 2 & Conv3-128 $\times$ 2 & $\left(
			\begin{tabular}{c}
			1 $\times$ 1, 128 \\
			3 $\times$ 3, 128 \\
			1 $\times$ 1, 512 
			\end{tabular}
			\right) \times 3$  \\ [3ex]
			\hline
			Conv3-256 $\times$ 3 & Conv3-256 $\times$ 4 & 	$\left(
			\begin{tabular}{c}
			1 $\times$ 1, 256 \\
			3 $\times$ 3, 256 \\
			1 $\times$ 1, 1024 
			\end{tabular}
			\right) \times 6$ \\ [3ex]
			\hline
			Conv3-512 $\times$ 6 & Conv3-512 $\times$ 8 & 	$\left(
			\begin{tabular}{c}
			1 $\times$ 1, 512 \\
			3 $\times$ 3, 512 \\
			1 $\times$ 1, 2048 
			\end{tabular}
			\right) \times 3$  \\ [3ex]
			\hline
			Max-Pooling & Max-Pooling & Average-Pooling  \\ [1ex] 
			\hline
			FC-4096 $\times$ 2 & FC-4096 $\times$ 2 &  \  \\ [1ex] 
			\hline
			FC-22 Softmax & FC-22  Softmax &  FC-22 Softmax  \\ [1ex] 
			\hline
		\end{tabular}\\
	
	\end{table}
	\normalsize 
		{\normalsize \noindent VGG16, VGG19 and ResNet50 architectures are shown. The ResNet50 model employs residual (skip) connections between the modules. The final layers of each of the networks were replaced with a new fully connected (FC) layer with 22 neurons followed by the softmax function.}
		
\begin{figure*}
		\centering     
		{\includegraphics[width=.75\textwidth]{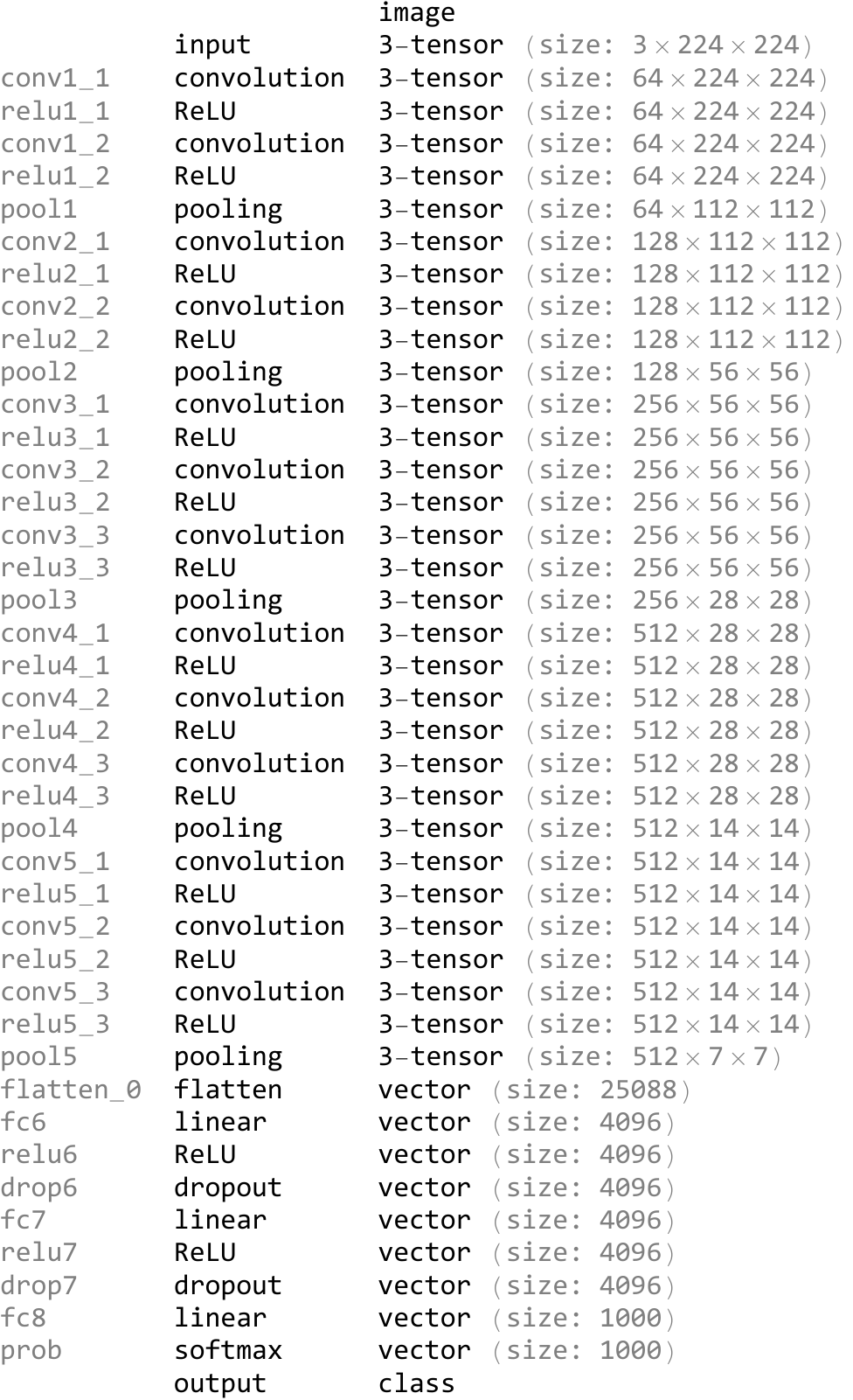}}		
		\caption{Architecture of VGG16. VGG19 has a similar linear design with additional convolution modules as shown in~\Tref{nets}.}
		\label{fig:VGG16}
	\end{figure*}

	\clearpage
	\noindent \Fref{fig:VGG16} presents the architecture of VGG16, and outlines the specific differences with VGG19.

	\clearpage

\begin{figure*}
		\centering     
		{\includegraphics[width=1.05\textwidth]{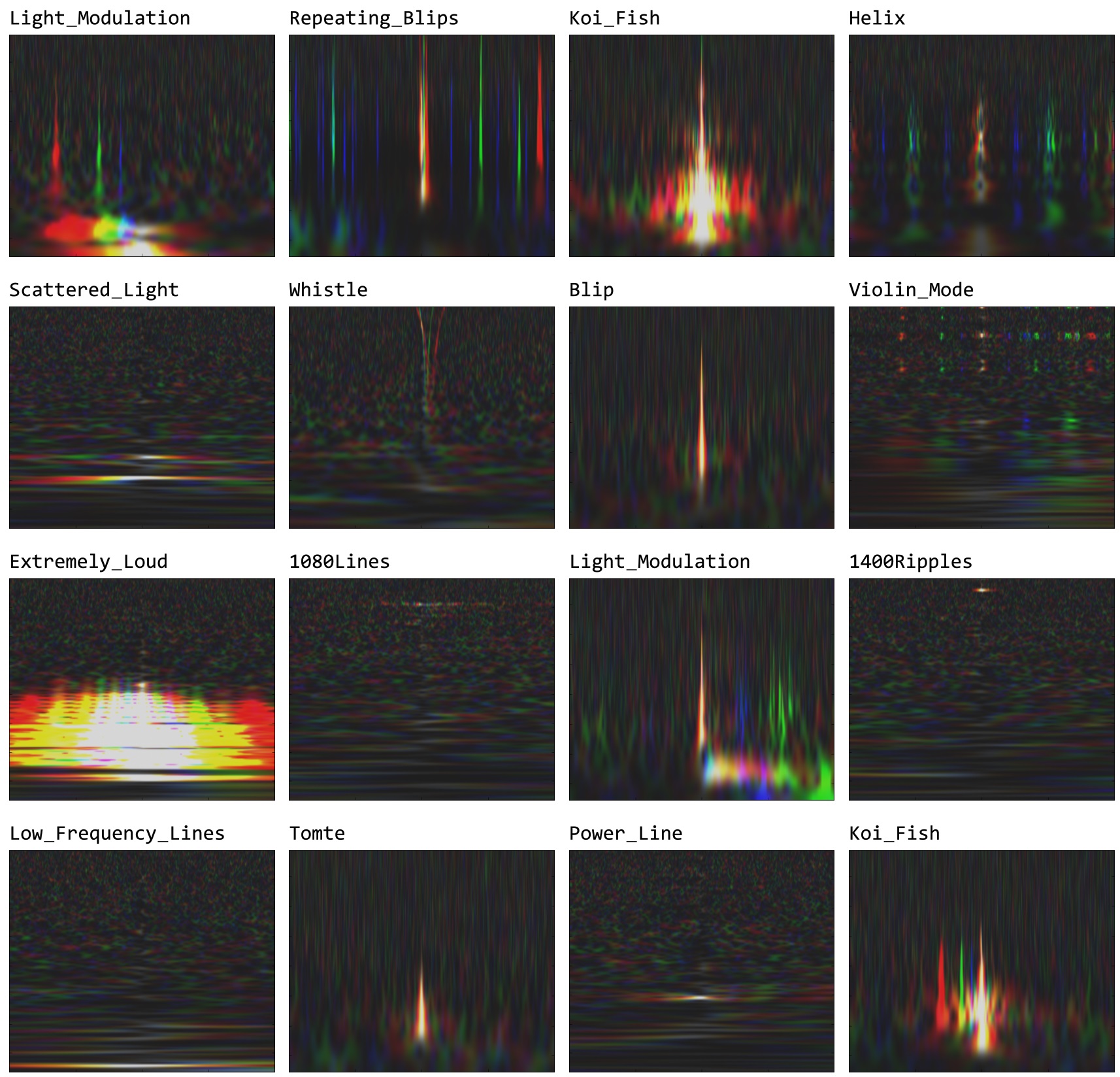}}		
		\caption{Random samples of multi-duration spectrograms encoded in single images. The RGB color channels correspond to 1.0s, 2.0s, and 3.0s durations respectively. This encoding was used to train and test the Inception models.}
		\label{fig:color}
	\end{figure*}

\begin{figure*}
		\centering     
		{\includegraphics[width=1.05\textwidth]{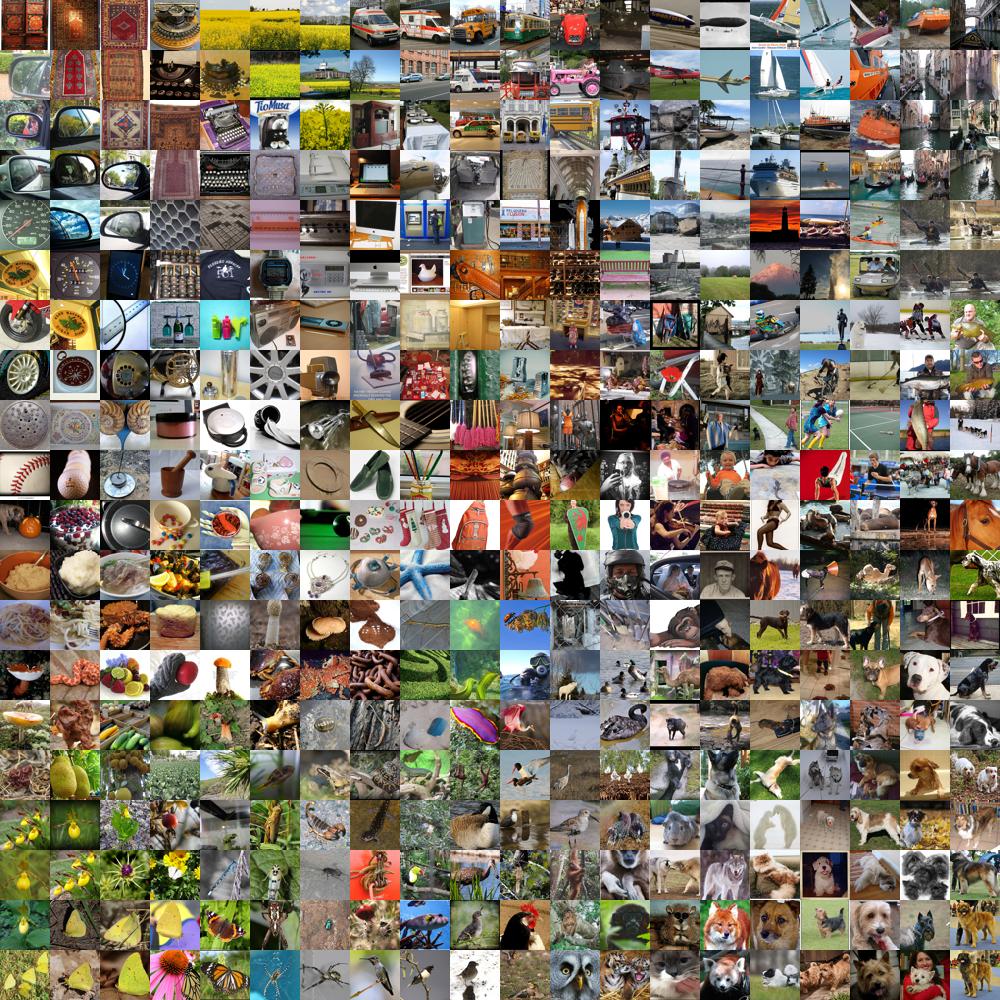}}		
		\caption{Random samples of images from the ImageNet database, which is a common benchmark used to test computer vision algorithms. It contains images of real-world objects belonging to 1000 different classes. Examples of classes include \texttt{hamster}, \texttt{taxi cab}, \texttt{stingray}, \texttt{tricycle}, \texttt{volleyball},  \texttt{lamp} ,\texttt{mushroom}, \texttt{restaurant}, \texttt{water bottle},  \texttt{hook}, \texttt{ladle}, \texttt{kite}, \texttt{speedboat}, etc. This dataset was used to pre-train the models used in this article for glitch classification.}
		\label{fig:image}
	\end{figure*}

	\clearpage
	\noindent \Fref{fig:color} presents samples of \texttt{Gravity Spy} glitches used to train and test our Inception models. \Fref{fig:image}	presents a gallery of samples from ImageNet. This database is used in the ILSVRC competitions to benchmark computer vision algorithms, and was used to pre-train the CNN models which we utilized for glitch classification in this article. 
	
	\section{}
	
	\begin{figure*}
		\centering     
		{\includegraphics[width=.6\textwidth]{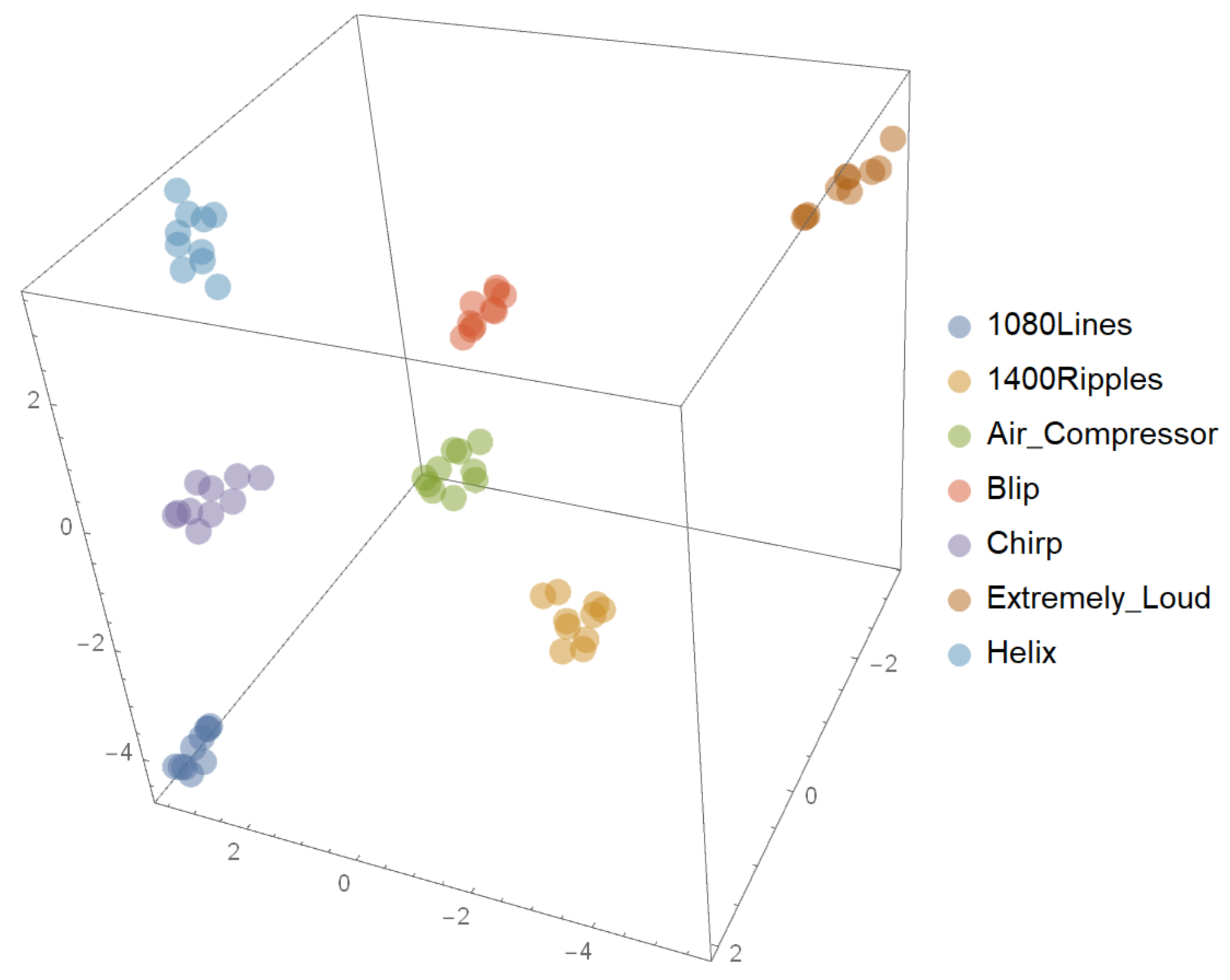}}		
		\caption{Location of different classes of glitches from the test set after applying the CNN feature extractor. The t-SNE algorithm was used to reduce the dimension from a 1008 vector to a 3D vector for visualization. Note that each type of glitch forms a cluster, and their relative positions depend on their morphology. Outliers may be inspected closely to verify whether the labels are accurate or whether they should belong to a new class.}
		\label{fig:tSNE1}
	\end{figure*}
	
	\begin{figure*}
		\centering     
		{\includegraphics[width=.7\textwidth]{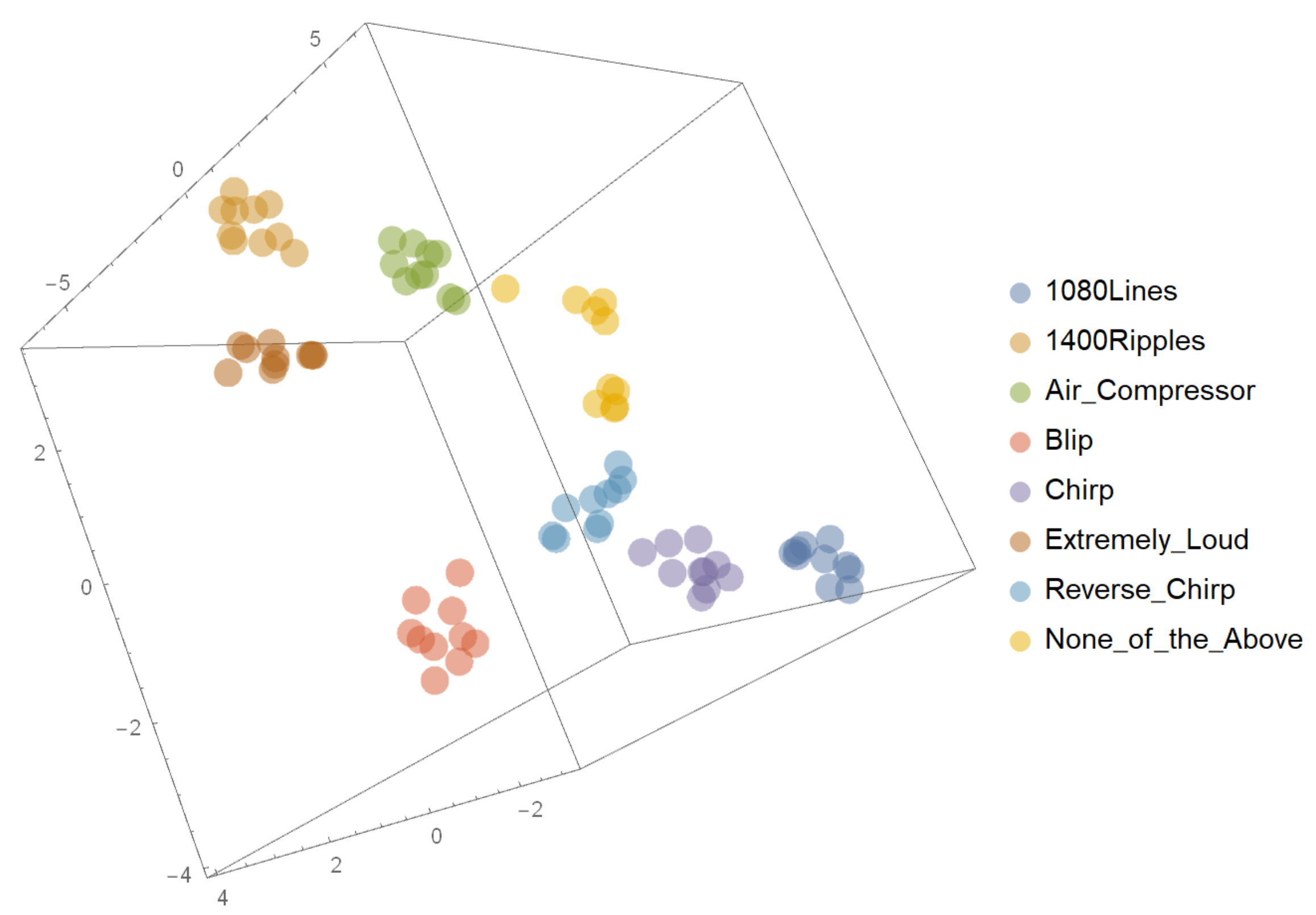}}		
		\caption{This is the CNN feature-space projected to 3D using t-SNE. A new class called \texttt{Reverse\_Chirp}, which are chirps reflected about the vertical axis, has been added. It can been seen that the CNN feature-extractor maps this class (which it has never seen before during training) to a unique cluster. Furthermore, this cluster is located near the Chirp class and the \texttt{None\_of\_the\_Above} class, which indicates that the relative positions of the glitches in this feature-space is meaningful and depends on their morphology.}
		\label{fig:tSNE2}
	\end{figure*}
	
	\begin{figure*}
		\centering     
		{\includegraphics[width=1.1\textwidth]{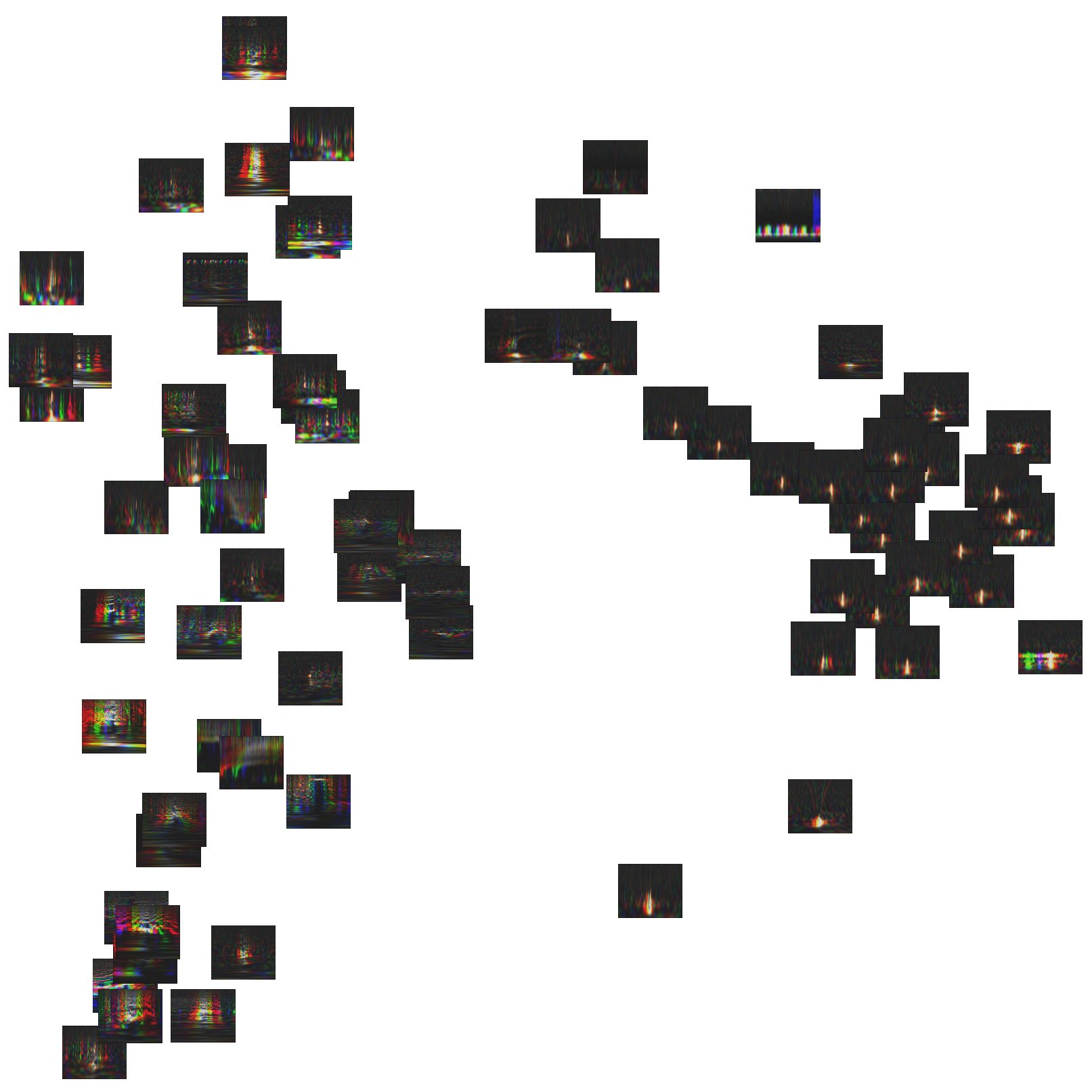}}		
		\caption{Unsupervised clustering of glitches belonging to none of the known classes. Note that glitches having similar morphology are located near each other. Our trained CNNs were used as feature extractors to map the images to a 1008-dimensional vector-space, which in turn was mapped to the 2-D space shown above using the t-SNE dimension reduction method. This feature-extractor can be used to group new classes of glitches by morphology and find the time of their occurrence in the future.}
		\label{fig:tSNE}
	\end{figure*}
	
	\begin{figure*}
		\centering     
		{\includegraphics[width=1.1\textwidth]{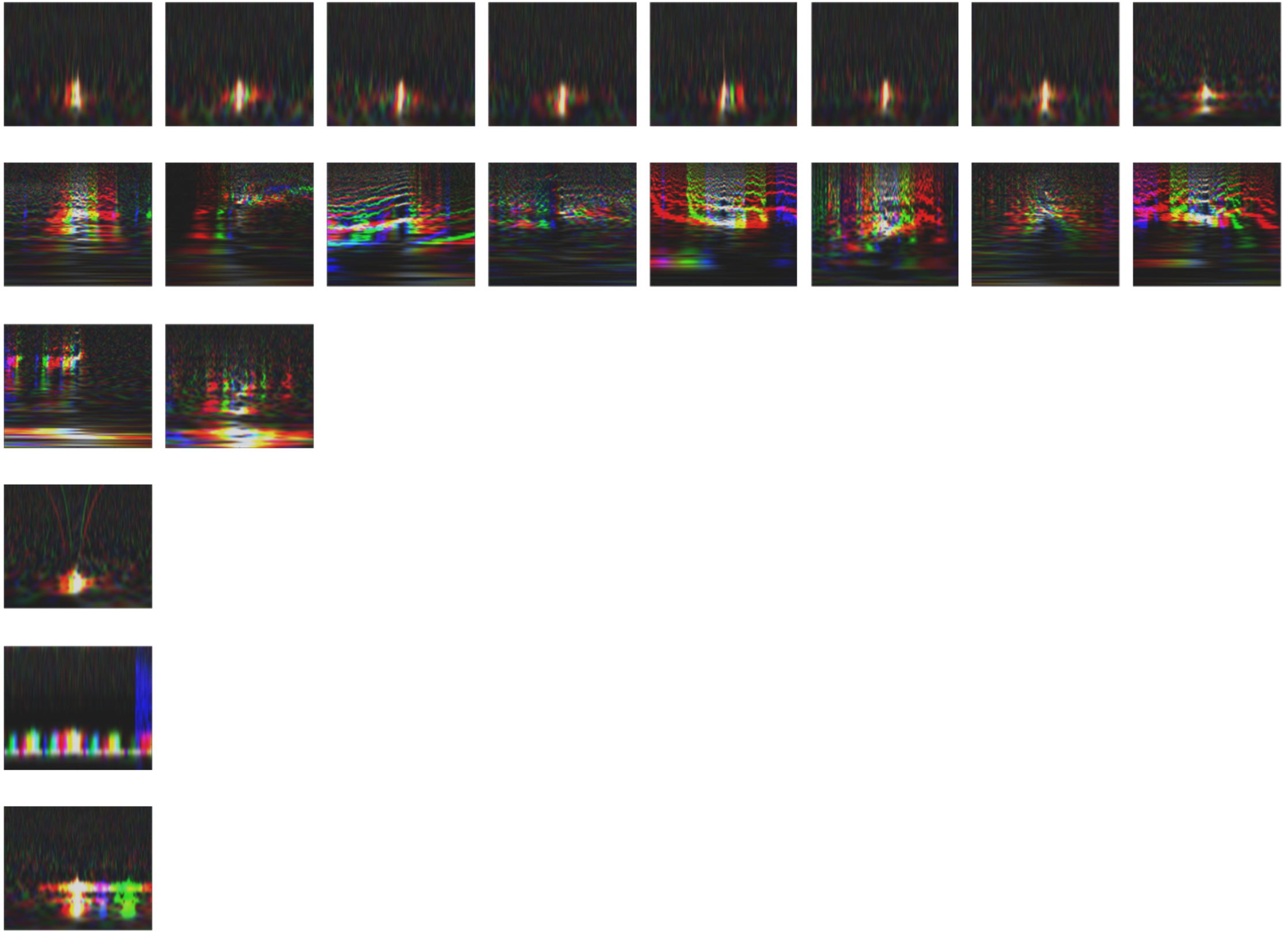}}		
		\caption{Some clusters of glitches in the \texttt{None\_of\_the\_Above} class automatically found by an unsupervised spectral clustering algorithm applied to the CNN feature-space shown in the previous Figure. Each row represents a cluster. Different clustering algorithms may be used, and their parameters can be tuned to vary the sensitivity of clustering to enforce stricter similarity criteria.}
		\label{fig:tSNE}
	\end{figure*}

\end{document}